\documentclass[aip,jcp,reprint]{revtex4-1}

\usepackage{natmove}
\usepackage{amsmath,amssymb}
\usepackage{graphicx}

\begin{document}
\title{Cross-stream migration of a Brownian droplet in a polymer solution under Poiseuille flow}

\author{Michael P. Howard}
\email{mphoward@utexas.edu}
\affiliation{McKetta Department of Chemical Engineering, University of Texas at Austin, Austin, Texas 78712, USA}

\author{Thomas M. Truskett}
\affiliation{McKetta Department of Chemical Engineering, University of Texas at Austin, Austin, Texas 78712, USA}

\author{Arash Nikoubashman}
\affiliation{Institute of Physics, Johannes Gutenberg University Mainz, Staudingerweg 7, 55128 Mainz, Germany}

\begin{abstract}
The migration of a Brownian fluid droplet in a parallel-plate microchannel was investigated using dissipative
particle dynamics computer simulations. In a Newtonian solvent, the droplet migrated toward the channel walls
due to inertial effects at the studied flow conditions, in agreement with theoretical predictions and recent simulations.
However, the droplet focused onto the channel centerline when polymer chains were added to the solvent. Focusing was
typically enhanced for longer polymers and higher polymer concentrations with a nontrivial flow-rate dependence due
to droplet and polymer deformability. Brownian motion caused the droplet position to fluctuate with a distribution
that primarily depended on the balance between inertial lift forces pushing the droplet outward and elastic forces
from the polymers driving it inward. The droplet shape was controlled by the local shear rate, and so its average
shape depended on the droplet distribution.
\end{abstract}
\maketitle

\section{Introduction}
Particle migration in a microchannel\cite{Stone:2001ws,Stone:2004ww} has important applications in separation technologies such as
filtration \cite{bhagat:physfluids:08}, cell sorting \cite{Hur:2011vz}, and fractionation \cite{Giddings:1993tq}. It also has implications for physical processes
like the margination of cells in the blood stream\cite{Kumar:2012ev,Kumar:2012en} and for multiphase flows in geological
formations (enhanced oil recovery) \cite{green:1998,wilson:1977,tehrani:jrheol:96}. Such cross-stream migration could be desirable if a separation is needed but
undesirable if a homogeneous distribution is preferred, and it is important to understand and design the
conditions under which migration occurs. Multiple mechanisms exist for cross-stream migration
in microchannels \cite{fu:natbiotech:99, cohen:prl:05, lee:appl:04, Das:2017eq,Das:2018dn}, but in this article we will focus on particle migration that is passively controlled by a
pressure- or gravity-driven flow \cite{karimi:biofluidics:13,amini:labchip:14}, which is attractive from an engineering perspective for its potential as a scalable,
high-throughput technology.

Rigid particles in a Newtonian fluid are known to move across streamlines in parabolic
(Poiseuille) flows due to lift forces at small but finite fluid inertia \cite{Leal:1980wy,Stone:2000js}. Inertial
lift outward from the channel center is balanced by an inward force induced by hydrodynamic interactions
with the walls, causing the particle to adopt an intermediate lateral position \cite{matas:jfluid:04, dicarlo:pnas:07, amini:labchip:14}. This effect was first
observed experimentally by Segr\'{e} and Silberberg \cite{segre:nature:61}, who found that millimeter-sized spheres in
pipe flow migrated to an annulus at roughly 60\% of the pipe radius. The number and position
of these ``focusing'' points depends on the channel geometry and flow, and has also been demonstrated for, e.g., parallel plates\cite{Ho:1974gm}
and square ducts \cite{dicarlo:prl:2009}.

Deformable droplets in a Newtonian fluid exhibit an even richer set of behaviors than their rigid counterparts \cite{Leal:1980wy}. Unlike
rigid spheres, droplets can migrate across streamlines even in the Stokes flow (inertialess) limit due to their deformability.
Chan and Leal showed that the direction of this migration depends on the viscosity ratio between the droplet and
the fluid \cite{Chan:1979ep}. Stan et al. found that chemical and surfactant-induced Marangoni effects also influenced droplet migration \cite{Stan:2011ez,Stan:2013cv}.
At finite fluid inertia, Legendre and Magnaudet demonstrated that there
is lift on a droplet \cite{Legendre:1997bh} analogous to the Saffman lift on a rigid particle \cite{Saffman:1965di,Saffman:1968jk} but with a magnitude that depends on the
viscosity ratio between the droplet and the fluid. Experiments \cite{Karnis:1966va,Hur:2011vz} and simulations \cite{Mortazavi:2000cv,Chen:2014ve,Pan:2016cd,Marson:2018fv} have shown that droplets
undergo Segr\'{e}--Silberberg-type migration in Poiseuille flow, and that the preferred lateral position
depends on several dimensionless parameters, as recently discussed in detail by Marson et al. \cite{Marson:2018fv}.

High-throughput applications like filtration or sorting may require focusing particles onto the channel centerline \cite{dicarlo:pnas:07,karimi:biofluidics:13},
which is not always achieved by inertial or deformation-induced migration in simple channel geometries.
Considerable efforts have been dedicated to design various microfluidic device geometries
that can manipulate particles in this way \cite{dicarlo:pnas:07}, but finding such geometries can be difficult and highly problem specific \cite{amini:labchip:14}.
Fortunately, it has been shown that the addition of polymers to the
Newtonian solvent provides a simple mechanism, called \textit{viscoelastic focusing} \cite{DAvino:2017kh}, to drive particles toward
regions of low shear.

Viscoelastic polymer solutions induce inward particle migration in Poiseuille flow due
to a gradient in the first normal stress difference over the particle surface \cite{leshansky:prl:07}. The elastic force exerted by the polymers
competes directly with other forces acting on the droplet for the flow conditions, including inertial lift, deformation-induced forces, and wall forces, to set the lateral position of the particle.
Such viscoelastic focusing of rigid particles has been demonstrated experimentally \cite{Karnis:1966tp, Gauthier:1971wr,Gauthier:1971vr, tehrani:jrheol:96, leshansky:prl:07, kim:labchip:12, Kim:2016tg}
and using computer simulations \cite{davino:labchip:2012,santo:prapp:2014,nikoubashman:jcp:14,nikoubashman:erratum:2014,Howard:2015bl,trofa:compfluids:2015}. Interestingly, a neutral surface separating focusing points at the channel center
and at the walls was discovered in simulations for certain classes of viscoelastic fluids \cite{davino:labchip:2012}.
Droplets under shear are also known to migrate in polymer solutions \cite{Hur:2011vz,Gauthier:1971wr,Gauthier:1971vr,Chan:1979ep}.

Most prior theoretical descriptions \cite{Chan:1979ep,Leal:1980wy,leshansky:prl:07,DAvino:2017kh} and simulations \cite{davino:labchip:2012,santo:prapp:2014,trofa:compfluids:2015} of viscoelastic focusing have adopted a
continuum-level description. Such models neglect microscopic details and fluctuations of the macromolecular
components of the viscoelastic medium and the particle or droplet.
However, in microfluidic and nanofluidic devices, it can be necessary to consider such motion and interactions.
For example, Brownian motion leads to considerable scattering in the position of a rigid sphere
around the Segr\'{e}--Silberberg annulus for Poiseuille flow in a pipe \cite{prohm:epje:12}. Moreover, Brownian particles
are often comparable in size to the macromolecular constituents of non-Newtonian fluids. At these length scales,
Brownian spheres can exhibit anomolous motion in polymer solutions \cite{Mackay:2003wq,Wong:2004cp,Tuteja:2007dz,PolingSkutvik:2015be}, which has been attributed to coupling between
the motion of the sphere and the polymers \cite{Chen:2018im,Chen:2018:2}. It is then unclear whether well-established results for viscoelastic
focusing of larger particles directly transfer to smaller particles in microchannels.

We previously demonstrated the applicability of viscoelastic focusing for Brownian rigid spheres with sizes
comparable to the constituent polymer chains of a viscoelastic medium \cite{nikoubashman:jcp:14,nikoubashman:erratum:2014,Howard:2015bl}. However, we noted significant fluctuations of
the particle around its focused position, in qualitative agreement with microfluidic experiments \cite{kim:labchip:12}.
It is desirable to exploit the viscoelastic focusing mechanism to manipulate small Brownian droplets,
which fluctuate in shape in addition to position, in microchannels. To our knowledge, this problem has
gone relatively unexplored. We hypothesize that similar considerations may apply to the droplets as
for the rigid spheres: namely, focusing onto the centerline should be improved by longer polymers and higher
polymers concentrations. However, as for the rigid spheres, the distribution of the droplet
position in the channel may be broad or narrow under certain flow conditions.

In this article, we test and confirm this hypothesis using particle-based computer simulations of the
cross-stream migration of a Brownian fluid droplet in a polymer solution under Poiseuille flow.
Although the droplet migrated outward in a Newtonian solvent (in agreement with prior simulations \cite{Marson:2018fv}),
we found that it focused onto the channel centerline in solutions of sufficiently long polymers at modest concentrations.
The flow-rate dependence of the focusing was nontrivial due to a combination of effects from droplet deformation and
the elastic force exerted by the polymers. We also varied the viscosity ratio between the droplet and the solvent,
but did not observe any significant effect on the viscoelastic focusing in the flow regime considered. The droplet shape was controlled
primarily by the local shear rate near the droplet, and so its average shape depended sensitively on the droplet distribution in the channel.

The rest of this article is organized as follows. We first describe the simulation model, including
characterization of the fluid surface tension and viscosity. We then report our results, first analyzing
the simulated flow fields and then systematically demonstrating the effects of polymer concentration,
polymer chain length, and flow rate on the distribution of the droplet in the channel and its shape.
We finally present our conclusions, suggesting avenues for future inquiry.

\section{Simulation model}
A single fluid droplet was simulated in a Newtonian solvent and in a polymer solution using dissipative particle dynamics (DPD)
simulations \cite{Hoogerbrugge:1992hl,Espanol:1995hf,Groot:1997du} DPD is a particle-based mesoscale simulation method that faithfully resolves hydrodynamic interactions,
incorporates thermal fluctuations, and is well-suited for modeling multiphase fluids.
In DPD, particles interact with each other through three pairwise forces: a conservative force $\mathbf{F}_{\rm C}$,
a dissipative force $\mathbf{F}_{\rm D}$, and a random force $\mathbf{F}_{\rm R}$. As is typical, we modeled
the conservative force acting on particle $i$ due to particle $j$ by a soft repulsion \cite{Groot:1997du},
\begin{equation}
\mathbf{F}_{\rm C} = \begin{cases}
a_{ij} \left(1-r/r_{\rm c}\right) \mathbf{\hat r} & r \le r_{\rm c} \\
0  & r > r_{\rm c}
\end{cases},
\end{equation}
where $a_{ij}$ sets the strength of the repulsion between particles $i$ and $j$, $r$ is the distance between
the particle centers, $\mathbf{\hat r}$ is the unit vector to the center of particle $i$ from the center of
particle $j$, and $r_{\rm c}$ is the cutoff radius for the interaction that sets the effective size of the particles.

The random and dissipative forces impart thermal fluctuations and drag while also acting as a thermostat on the DPD particles.
These forces are applied in a pairwise manner that conserves momentum, with the forces on particle $i$ from particle $j$ given by
\begin{align}
\mathbf{F}_{\rm D} &= -\gamma_{ij} w(r)(\mathbf{\hat r}\cdot \Delta \mathbf{v}) \mathbf{\hat r} \\
\mathbf{F}_{\rm R} &= \sqrt{\gamma_{ij} w(r)} \xi \mathbf{\hat r},
\end{align}
where $\gamma_{ij}$ is the drag coefficient between particles $i$ and $j$, $w$ is a weight function, and
$\Delta \mathbf{v} = \mathbf{v}_i - \mathbf{v}_j$ is the difference in the velocities of particles $i$ and $j$.
To satisfy the fluctuation--dissipation theorem \cite{Espanol:1995hf},
$\xi$ is an independent random variable for each pair of particles
that has zero mean, $\langle \xi(t) \rangle = 0$, and a variance $\langle \xi(t) \xi(t') \rangle = 2 k_{\rm B} T \delta(t-t')$
with $k_{\rm B}$ being Boltzmann's constant and $T$ being the temperature. In this work, the drag coefficients were assigned per particle, $\gamma_i$,
and the effective drag coefficient for a pair was determined by the mixing rule $\gamma_{ij} = 2/(1/\gamma_i + 1/\gamma_j)$ \cite{Visser:2006bk}.

The weight function $w$ modulates the dynamic properties, i.e., diffusivity and viscosity, of the fluid. We used
the generalized weight function proposed by Fan et al. \cite{Fan:2006ex},
\begin{equation}
w(r) = \begin{cases}
(1-r/r_{\rm c})^s & r \le r_{\rm c} \\
0  & r > r_{\rm c}
\end{cases},
\end{equation}
with $s = 1/2$. This choice of $s$ increases the Schmidt number of the fluid compared to the standard DPD weight
function \cite{Groot:1997du} ($s = 2$) to give a value closer to that of a real liquid. We also found that using $s=1/2$ gave
better agreement with the no-slip boundary conditions at the microchannel walls than using $s=2$ (see below).

\subsection{Fluid model}
The polymer solution and droplet were modeled using three types of DPD particles:
solvent (s) particles, polymer segment (p) particles, and droplet (d) particles.
The model and results in this article will be reported in a fundamental system of units using $d$ as the
unit of length, $m$ as the unit of mass, and $\varepsilon$ as the unit of energy, which gives
$\tau = \sqrt{m d^2/\varepsilon}$ as the unit of time. Throughout, the total density of
DPD particles was $\rho = 3.0/d^3$, all DPD particles had equal mass $1.0\,m$, the temperature was
$T = 1.0\,\varepsilon/k_{\rm B}$, and the cutoff radius was $r_{\rm c} = 1.0\,d$.
All simulations were performed using HOOMD-blue \cite{Anderson:2008vg,Glaser:2015cu,Phillips:2011td} (version 2.2.5) on multiple graphics processing units with a
simulation time step of $0.01\,\tau$.

In order to choose the DPD repulsive parameters, we first computed the surface tension $\sigma$ between coexisting
slabs of solvent and droplet particles. We fixed the repulsive parameter for particles of the same type to
standard DPD values \cite{Groot:1997du}, $a_{\rm ss} = a_{\rm dd} = 25\,\varepsilon/d$, but varied the cross-interaction strength, $a_{\rm sd}$.
The drag coefficient should not affect the measured surface tension, which is a static property, and so was fixed
to $\gamma_{\rm s} = \gamma_{\rm d} = 1.0\,m/\tau$ to promote fast diffusion. The coexisting slabs were equilibrated by
joining two cubic regions of edge length $30\,d$ to give an orthorhombic box centered around the origin with edge lengths
$L_x = 30\,d$, $L_y = 30\,d$, and $L_z = 60\,d$, where $x$, $y$, and $z$ denote the Cartesian coordinate axes.
Particles were allowed to interdiffuse for $5 \times 10^4\,\tau$ to equilibrate the joined slabs.

\begin{figure}[h]
    \centering
    \includegraphics{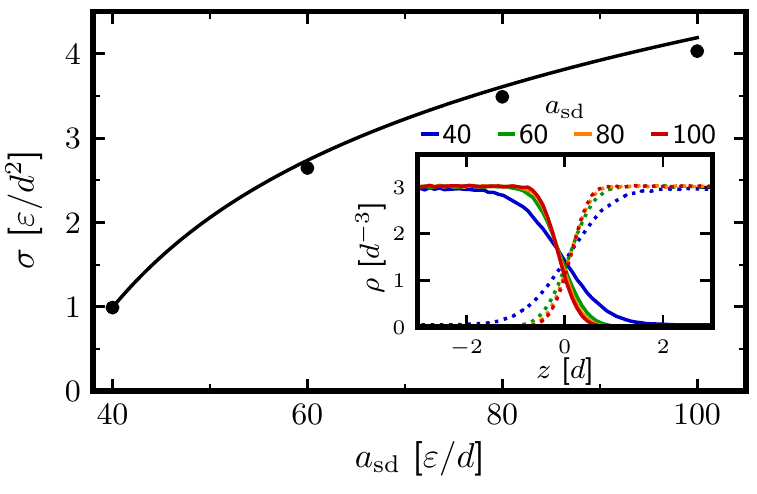}
    \caption{Surface tension $\sigma$ between slabs of solvent and droplet particles for varied strengths of the repulsive
             cross-interaction $a_{\rm sd}$. The solid line gives the value predicted by eq.~\eqref{eq:groot}. Inset:
             Density $\rho$ of solvent (solid) and droplet (dashed) particles near the fluid interface at $z = 0\,d$.}
    \label{fgr:surftens}
\end{figure}

In this geometry, $\sigma$ can be computed from the pressure anisotropy \cite{Kirkwood:1949gf,frenkel},
\begin{equation}
\sigma = \frac{L_z}{2} \left\langle p_{zz} - \frac{p_{xx} + p_{yy}}{2} \right\rangle,\label{eq:surftens}
\end{equation}
where $p_{\alpha\alpha}$ denotes the diagonal component of the stress tensor for index $\alpha$, and the prefactor of $1/2$ accounts for the presence of
two interfaces due to the periodic boundary conditions.
The cross-interaction strength was varied from $a_{\rm sd} = 40\,\varepsilon/d$ to $100\,\varepsilon/d$,
and the surface tension was measured using eq.~\eqref{eq:surftens} by sampling $p_{\alpha\alpha}$ every $0.05\,\tau$
during a $10^5\,\tau$ simulation.
The measured surface tension (Fig.~\ref{fgr:surftens}) is in good agreement with Groot and Warren's empirical equation \cite{Groot:1997du},
\begin{equation}
\sigma = 0.75 \rho k_{\rm B} T r_{\rm c} \chi^{0.26} (1-2.36/\chi)^{3/2}, \label{eq:groot}
\end{equation}
with $\chi = 0.286(a_{\rm sd}-a_{\rm ss})$ being their fit to the Flory--Huggins parameter for $\rho = 3.0/d^3$.

As expected, the surface tension increased with increasing $a_{\rm sd}$
because the solvent and droplet particles became less miscible. The solvent and droplet particle density profiles near the
interface (inset of Fig.~\ref{fgr:surftens}) converged to similar values with increasing $a_{\rm sd}$.
We desired a droplet that was sparingly soluble in the solvent but that could still deform under the flow rates
accessible in the simulations. At $a_{\rm sd} = 60\,\varepsilon/d$, the density of solvent particles dissolved in the droplet phase was already
small ($4.66\times 10^{-4}/d^3$), and the surface tension $\sigma = 2.65\,\varepsilon/d^2$ permitted modest deformation
under viable flow rates.
We accordingly selected $a_{\rm sd} = 60\,\varepsilon/d$ for the cross-interaction strength.

We subsequently measured the shear viscosity of the solvent using reverse nonequilibrium simulations (RNES) \cite{MullerPlathe:1999vc}.
Details of this method are well-described elsewhere \cite{MullerPlathe:1999vc,Statt:2018}. We simulated a cubic box of edge length $20\,d$ containing
only solvent particles with drag coefficients that varied from $\gamma_{\rm s} = 1.0\,m/\tau$ to $50.0\,m/\tau$.
Using RNES, we imposed a shear stress $\tau_{zx}$ on the solvent by periodically exchanging the $x$-momenta of
one pair of particles from slabs of width $1.0\,d$ centered at $z = \pm 5\,d$. The swapped particles were the
ones that most opposed the desired direction of flow ($x$) in each slab. We measured the velocity profile $u_x(z)$ between
the exchange slabs ($|z| < 3.5\,d$) every $10\,\tau$ over a $5 \times 10^5\,\tau$ simulation, obtaining a
Couette flow profile with a shear rate $\dot{\gamma} = \partial u_x / \partial z$ that decreased as the time between exchanges was increased
from $0.05\,\tau$ to $0.5\,\tau$. The imposed shear stress was proportional to the measured shear rate,
$\tau_{zx} = \mu_{\rm s} \dot{\gamma}$, as expected for a Newtonian fluid.
The shear viscosity, $\mu_{\rm s}$, was then determined by a linear fit of $\tau_{zx}$ versus $\dot{\gamma}$.

\begin{figure}[h]
    \centering
    \includegraphics{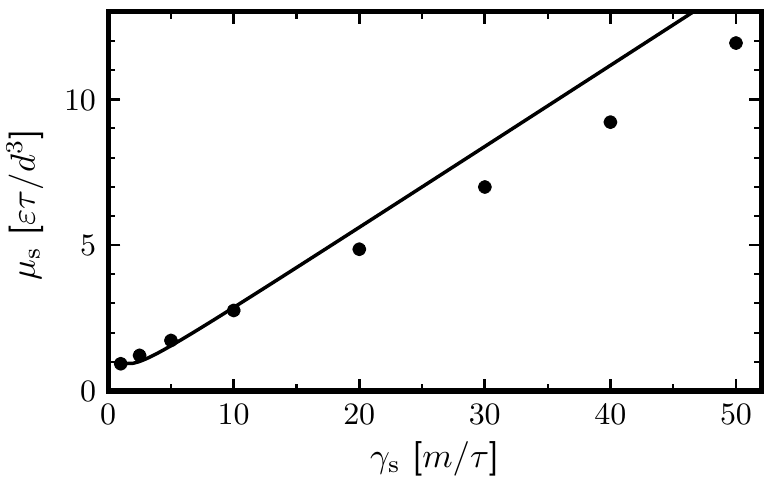}
    \caption{Solvent viscosity $\mu_{\rm s}$ for varied drag coefficient $\gamma_{\rm s}$. The solid line gives the
             value predicted by eq.~\eqref{eq:fan}.}
    \label{fgr:viscosity}
\end{figure}

As expected, the viscosity increased with increasing $\gamma_{\rm s}$ (Fig.~\ref{fgr:viscosity}). The simulated viscosity
was generally lower than theoretically estimated \cite{Fan:2006ex},
\begin{equation}
\mu_{\rm s} = \frac{315 k_{\rm B} T}{128\pi\gamma_{\rm s} r_{\rm c}^3} + \frac{512\pi \gamma_{\rm s} \rho^2 r_{\rm c}^5}{51975}, \label{eq:fan}
\end{equation}
particularly at high values of $\gamma_{\rm s}$.
We selected $\gamma_{\rm s} = 4.5\,m/\tau$ for the solvent particles, which gives a measured shear viscosity of
$\mu_{\rm s} = 1.73 \varepsilon \tau/d^3$. In most cases, we used $\gamma_{\rm d} = \gamma_{\rm s}$, giving a droplet
viscosity $\mu_{\rm d} = \mu_{\rm s}$, but we also varied $\gamma_{\rm d}$ to determine the effects of the viscosity ratio
$\mu_{\rm d}/\mu_{\rm s}$ in select cases.

To model linear polymer chains of length $M$ that were fully soluble in the solvent but insoluble in the droplet, the
polymer segment (p) particles were treated as if they were solvent (s) particles in the DPD interactions,
i.e., $a_{\rm pp} = a_{\rm ps} = 25\,\varepsilon/d$, $a_{\rm pd} = 60\,\varepsilon/d$, and $\gamma_{\rm p} = 4.5\,m/\tau$.
Bonds within a chain were modeled by adding a harmonic spring force $\mathbf{F}_{\rm B}$ to $\mathbf{F}_{\rm C}$ for
connected pairs of particles. The force on particle $i$ bonded to particle $j$ was
\begin{equation}
\mathbf{F}_{\rm B} = -\kappa \left(r - b \right) \mathbf{\hat r},
\end{equation}
with spring constant $\kappa = 100\,\varepsilon/d^2$ and $b = 0.7\,d$ \cite{Kranenburg:2004fu}.

\subsection{Flow in microchannel}
We simulated gravity-driven flow of the droplet and polymer solution in a parallel plate microchannel.
The full system was initialized as follows.
We first dispersed solvent particles with a total density of $\rho = 3.0/d^3$ into a three-dimensional, periodic
simulation box of dimensions $L_x = 80\,d$, $L_y = 40\,d$, and $L_z = 42\,d$ and equilibrated the solvent for $1000\,\tau$.
We chose $x$ as the direction of flow in the microchannel, and the parallel plates had normals along $z$.
We constructed the microchannel walls by freezing solvent particles having $|z| \ge H = 20\,d$, zeroing their velocities,
and switching their types to be wall (w) particles \cite{Pivkin:2005bu,Fedosov:2008ct}. (The total channel width was $2H$.)
The wall particles interacted with the fluid as if they were
solvent particles, i.e., $a_{\rm sw} = a_{\rm pw} = 25\,\varepsilon/d$, $a_{\rm dw} = 60\,\varepsilon/d$,
and $\gamma_{\rm w} = 4.5\,m/\tau$. Mutual DPD interactions between wall particles were excluded.
To help enforce no-slip and no-penetration boundary conditions at the walls,
solvent, polymer, and droplet particles were additionally reflected from the planes at $z = \pm H$ using
bounce-back rules \cite{Fedosov:2008ct,Revenga:1999wl}.

We selected particles near the origin of the channel to form a droplet of radius $R = 4.0\,d$,
giving a droplet blockage ratio of $R/H \approx 0.2$.
Due to the sparing solubility of the droplet particles in the solvent, we first estimated the number
of particles required to form such a droplet volume using the lever rule with the coexistence densities
shown in Fig.~\ref{fgr:surftens}. This procedure gave a droplet with a radius initially larger than $R$, but
some particles later dissolved into the solvent so that the droplet reached its target radius.
We then randomly created linear polymers of length $M$ from the remaining solvent particles.
To build each chain, we first randomly removed $M$ solvent particles. They were reinserted
as polymer segment (p) particles between the channel walls in a randomly generated chain conformation
having a bond length of $0.7\,d$ between connected particles. The number of polymer chains $N_{\rm p}$ was chosen to give the desired polymer weight
fraction, $\phi_{\rm p} = N_{\rm p} M/\rho V$, where $V = 2 L_x L_y H$ is the volume of the microchannel.
In most simulations, we used $\phi_{\rm p} = 5.0\%$ or $10.0\%$, but also tested $\phi_{\rm p} = 0.0\%$ (no polymer),
$2.5\%$ and $7.5\%$ for selected conditions.
The complete configuration, including the solvent, droplet, and polymers, was equilibrated for $5000\,\tau$.

\begin{figure}[h]
    \centering
    \includegraphics[width=8cm]{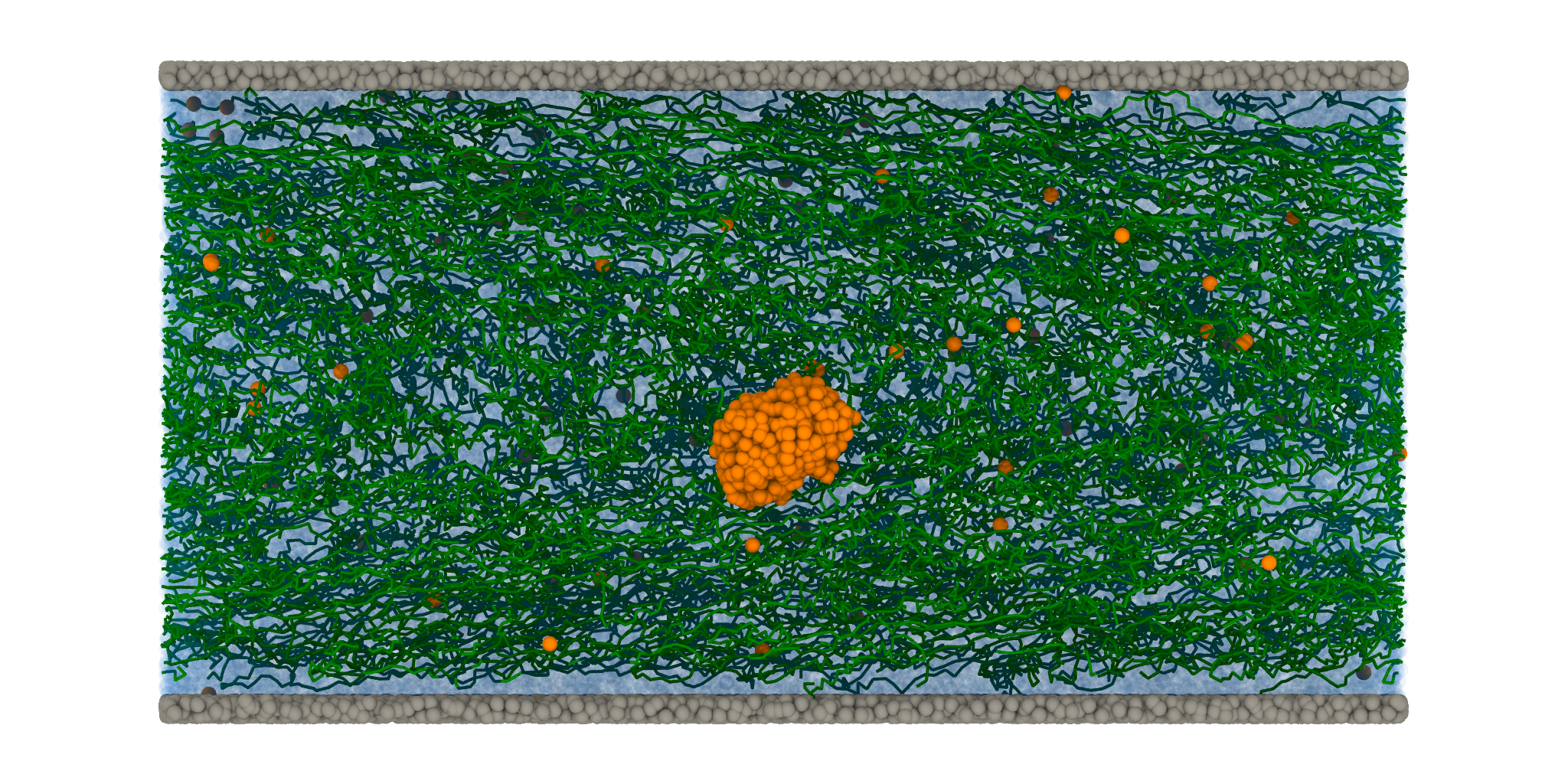}
    \caption{Fluid droplet (orange) in a parallel plate microchannel (gray) with $f_x = 0.005\,\varepsilon/d$.
             Polymers of length $M=80$ at polymer weight fraction $\phi_{\rm p} = 10\%$ are depicted in green.
             The solvent particles (blue) have been removed from the front of the image for visual clarity.
             This snapshot was rendered using OVITO 2.9.0 \cite{Stukowski:2010ky}.}
    \label{fgr:snapshot}
\end{figure}

Flow was generated by applying a constant body force, $f_{\rm x}$, in the $x$-direction for all solvent, polymer,
and droplet particles. For the pure solvent, applying such a force in conjunction with no-slip boundary
conditions at the channel walls gives the standard parabolic (Poiseuille) velocity field,
\begin{equation}
u_x(z) = U \left[ 1 - \left(\frac{z}{H}\right)^2 \right], \label{eq:velocity}
\end{equation}
where $U = \rho f_x H^2 / 2\mu_{\rm s}$ is the maximum velocity at the channel centerline for this flow field.
To help enforce the wall boundary conditions in the simulations, the frozen wall particles
were assigned velocities $v_x(z) = -u_x(2H - |z|)$ based on their positions in the wall \cite{Fedosov:2008ct}.
Additionally, $u_x(z)$ was initially superimposed onto the polymer solution and droplet to accelerate
the approach to a steady flow profile during a $5000\,\tau$ simulation. Fig.~\ref{fgr:snapshot}
shows an example configuration for the polymer solution under flow at steady state.

We repeated this procedure 5 times for each combination of chain length $M$, polymer concentration $\phi_{\rm p}$,
body force $f_{\rm x}$, and droplet viscosity $\mu_{\rm d}$ studied to generate independent starting configurations.
Production simulations of $10^5\tau$ were performed for each configuration. The droplet properties were sampled every $50\,\tau$, while the
properties of the entire solution were recorded every $2500\,\tau$. The computational workflow and data were managed using the signac framework \cite{signac_commat}.

\section{Results and discussion}
\subsection{Flow field}
\begin{figure}[h]
    \centering
    \includegraphics{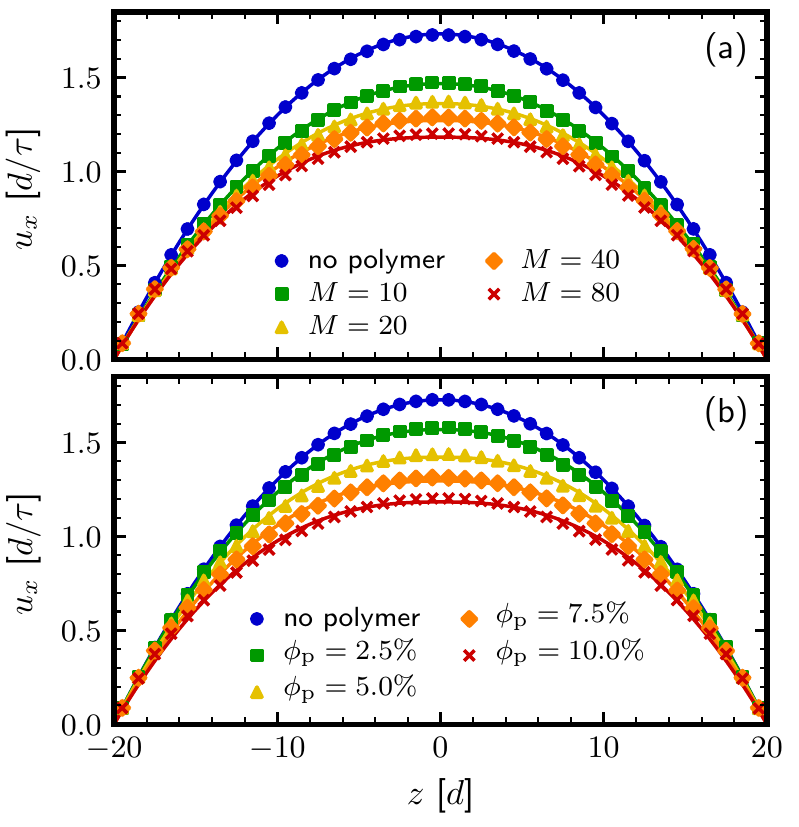}
    \caption{Average flow profile in the microchannel, $u_x$, at $f_x = 0.005\,\varepsilon/d$ for (a) various polymer chain lengths at $\phi_{\rm p} = 10.0\%$
             and (b) various concentrations of $M=80$ polymers. The solid lines give the expected profile according to eq.~\eqref{eq:velocity}
             without any fitting parameters for the no-polymer case and the fitted profiles from eq.~\eqref{eq:powerlaw}
             for the polymer solutions. Note that the circles and crosses display the same data in both panels.}
    \label{fgr:velocity}
\end{figure}

We first measured the average flow field in the microchannel, including the solvent, polymers, and droplet.
The flow was unidirectional, $u_x(z)$, and is shown for various polymer
chain lengths at the largest polymer concentration simulated ($\phi_{\rm p} = 10.0\%$) in Fig.~\ref{fgr:velocity}a
and for various concentrations of the longest polymers simulated ($M=80$) in Fig.~\ref{fgr:velocity}b.
The body force in Fig.~\ref{fgr:velocity} was $f_x = 0.005\,\varepsilon/d$, which was the largest value we
simulated and where any wall slip or non-Newtonian flow effects should be most pronounced.
This upper bound for $f_x$ in our simulations was determined by trial and error so that
no droplet breakup occurred.

The velocity profile in the absence of polymer was parabolic,
as expected, and was also in quantitative agreement with eq.~\eqref{eq:velocity}
using the measured $\mu_{\rm s}$ (Fig.~\ref{fgr:viscosity}). This indicates that the no-slip boundary conditions
are well-enforced and also validates the RNES measurement of $\mu_{\rm s}$.
The addition of polymers with $M=10$ resulted in a lower maximum velocity $U$ at the centerline,
consistent with the expected higher viscosity of a polymer solution (Fig.~\ref{fgr:velocity}a) \cite{Rubinstein}.
Increasing the length of the polymers from $M=10$ to $M=80$ further lowered $U$. Additionally,
the velocity profiles became less parabolic and developed a flattened region near $z=0$,
consistent with an increasingly non-Newtonian character of the fluid.
Similar trends were observed when varying the concentration of the $M=80$ polymers from $\phi_{\rm p} = 0.0\%$
to $10.0\%$ (Fig.~\ref{fgr:velocity}b), with higher polymer concentrations giving less parabolic flow profiles.

To quantify this non-Newtonian behavior, we modeled the shear stress in the polymer solutions using a
power law, $\tau_{zx} = K \dot \gamma^n$, where $n$ is the flow behavior index and $K$ is a prefactor giving correct dimensions to $\tau_{zx}$.
In a parallel plate channel, the flow field for a power-law fluid is
\begin{equation}
u_x(z) = \frac{n}{n+1} \left( \frac{\rho f_x H^{n+1}}{K} \right)^{1/n}\left[1 - \left(\frac{|z|}{H}\right)^{1+1/n} \right]. \label{eq:powerlaw}
\end{equation}
A Newtonian solvent has $n = 1$ and $K = \mu_{\rm s}$, and eq.~\eqref{eq:powerlaw} reduces to eq.~\eqref{eq:velocity}, whereas shear-thinning
fluids have $n < 1$. We determined $K$ and $n$ by fitting the flow fields in Fig.~\ref{fgr:velocity} through
eq.~\eqref{eq:powerlaw}, recovering exponents ranging
from $n = 0.92$ for $M = 10$ to $n = 0.66$ for $M = 80$ when $\phi_{\rm p} = 10.0\%$ (Fig.~\ref{fgr:velocity}a).
Likewise, $n$ decreased from $0.81$ at $\phi_{\rm p} = 2.5\%$ to $0.69$ at $\phi_{\rm p} = 7.5\%$
for the $M=80$ polymers (Fig.~\ref{fgr:velocity}b).
There is a small but noticeable deviation of the measured velocity from the fit using eq.~\eqref{eq:powerlaw} for $M=80$
at $\phi_{\rm p} = 10.0\%$, suggesting that the shear stress may have a more sophisticated functional form than the power-law model.
Nonetheless, the fitted exponents give us a useful qualitative characterization of the polymer solutions.

Longer polymer chains shear thin more readily under flow than shorter chains because they have
longer relaxation times, $\tau_{\rm p}$, that cause them to deform and align with the flow at smaller $\dot{\gamma}$ \cite{doi}.
The dimensionless Weissenberg number, ${\rm Wi} = \dot{\gamma} \tau_{\rm p}$, characterizes this relationship.
When ${\rm Wi} \ll 1$, the rate of deformation is slow compared to the polymer relaxation
and primarily coil conformations are expected, whereas for ${\rm Wi} \gg 1$, the polymers are expected to be highly deformed.
We approximate the shear rate by $\dot{\gamma} \approx U/H$, and estimate the polymer relaxation time
from the Zimm model for a Gaussian polymer chain in dilute solution \cite{doi},
$\tau_{\rm p} \approx \mu_{\rm s} b^3 M^{3/2} / k_{\rm B}T$.
We find that ${\rm Wi} \approx 1.3$ for the $M = 10$ polymers and
${\rm Wi} \approx 24$ for the $M = 80$ polymers for the conditions in Fig.~\ref{fgr:velocity}a ($f_x = 0.005\,\varepsilon/d$).
Hence, more significant shear-alignment is expected for the
longer chains, which should result in more shear thinning (smaller values of $n$). This expectation
is consistent with the shape of the flow fields in Fig.~\ref{fgr:velocity} and the fitted values for $n$.

\subsection{Droplet distribution}
Having characterized the flow in the microchannel, we measured the center-of-mass position of the droplet
between the channel walls, $z_{\rm c}$. The droplet was identified for each configuration using a clustering
procedure \cite{scikit-learn,Ester:1999tm} in order to exclude droplet particles dissolved in the solvent from subsequent analysis.
We analyzed the absolute value $|z_{\rm c}|$ based on the symmetry of the microchannel and to improve sampling.
Previous studies \cite{Marson:2018fv,Pan:2016cd} have reported the average center-of-mass position, $\langle |z_{\rm c}| \rangle$, which
is the first moment of the distribution of $|z_{\rm c}|$. However, a Brownian droplet can adopt a variety of
distributions in the channel depending on the conditions, and we found that $\langle |z_{\rm c}| \rangle$
was not sufficiently discriminating between these.
For example, a uniformly distributed droplet has $\langle |z_{\rm c}| \rangle \approx (H-R)/2$, which is
indistinguishable from a droplet which is strongly focused at this position throughout the entire simulation.
We accordingly computed the distribution of $|z_{\rm c}|$ using a bin size of $1.0\,d$,
and will focus most of our discussion around such distributions.

\begin{figure}[h]
    \centering
    \includegraphics{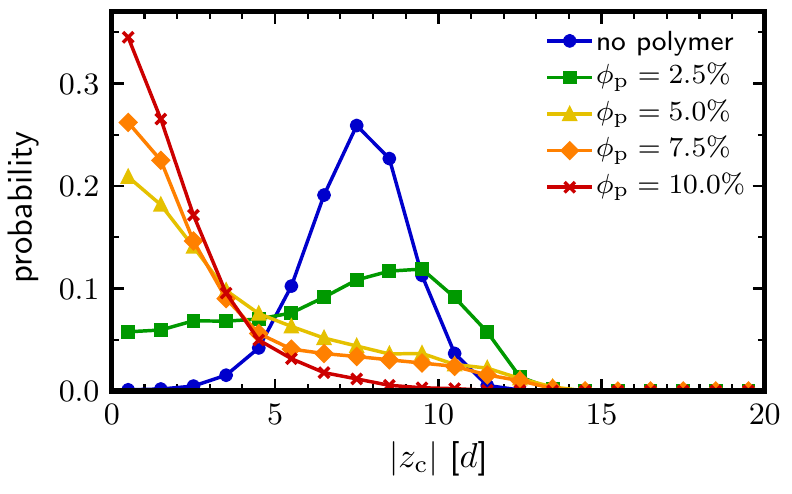}
    \caption{Distribution of droplet center-of-mass, $|z_{\rm c}|$, in a solution of $M=80$ polymers at increasing
             polymer concentrations $\phi_{\rm p}$ for $f_x = 0.005\,\varepsilon/d$.}
    \label{fgr:distconc}
\end{figure}

We first considered the distribution of the droplet in solutions of $M = 80$ polymer chains of increasing
concentration $\phi_{\rm p}$ (Fig.~\ref{fgr:distconc}) for the flow conditions shown in Fig.~\ref{fgr:velocity}b.
In the neat solvent, the droplet migrated outward from the channel centerline, showing a strongly preferred
position of $|z_{\rm c}| \approx 7.5\,d$. Such outward migration is consistent with prior theoretical and simulation work for droplets \cite{Chen:2014ve,Marson:2018fv,Pan:2016cd}.
The droplet can migrate by two mechanisms: (1) deformation due to the flow, even in the creeping flow limit,
and (2) lift forces at finite inertia. We define a channel Reynolds number, ${\rm Re} = 2 \rho U H/\mu_{\rm s}$,
and a droplet Reynolds number ${\rm Re}_{\rm d} = {\rm Re} (R/H)^2$ \cite{dicarlo:pnas:07,Marson:2018fv}. When the channel Reynolds number is sufficiently
small, the flow is expected to be laminar. When ${\rm Re}_{\rm d}$ is small, inertial forces on the droplet
are not significant and results from the Stokes flow limit are expected to apply. As an upper bound, we find
${\rm Re} \lesssim 100$ and ${\rm Re}_{\rm d} \lesssim 4$ for the investigated flow rates, consistent with the laminar flow of Fig.~\ref{fgr:velocity} but suggesting
that inertial lift on the particle may be significant. This estimate is in accord with the observed migration of the
droplet away from the centerline.

The addition of polymers to the channel at increasing polymer concentration $\phi_{\rm p}$ dramatically altered the
preferred position of the droplet. The droplet distribution significantly
broadened at the lowest concentration ($\phi_{\rm p} = 2.5\%$).
Interestingly, this included an increased probability of finding the droplet near the wall, beyond the preferred
peak in the neat solvent, which we speculate may be partially due to polymer-mediated depletion interactions \cite{Asakura:1954ts}.
Depletion, often discussed in the context of rigid spherical colloids in solution with
smaller polymer chains, induces an effective attraction between otherwise hard particles (the colloids) due to volume
exclusion of a secondary species (the polymers) \cite{Russel:1989}. In this case, the effective attraction is between the droplet and the wall because the polymers are
insoluble in the droplet and cannot penetrate the channel boundaries. Such an attraction near the wall was more
pronounced for simulations at smaller $f_x$ (not shown here).

Continuing to add polymer increasingly focused the droplet onto the channel centerline with a narrowing
distribution of $|z_{\rm c}|$. The increased polymer concentration had competing effects on the droplet migration.
On the one hand, the depletion force scales with $\phi_{\rm p}$ \cite{Asakura:1954ts}, which increases the outward force
on the droplet near the wall. On the other hand, the increased $\phi_{\rm p}$ lowered the maximum velocity
in the channel, decreasing the outward inertial lift on the droplet \cite{dicarlo:pnas:07}. Concurrently, the increased polymer
concentration also increased the inward elastic force on the droplet \cite{Howard:2015bl}.
The net result of these interactions is an increased inward force for larger $\phi_{\rm p}$,
which improves the droplet focusing onto the centerline, consistent with
with our previous work on viscoelastic focusing of rigid particle \cite{nikoubashman:jcp:14,nikoubashman:erratum:2014,Howard:2015bl}.

\begin{figure}[h]
    \centering
    \includegraphics{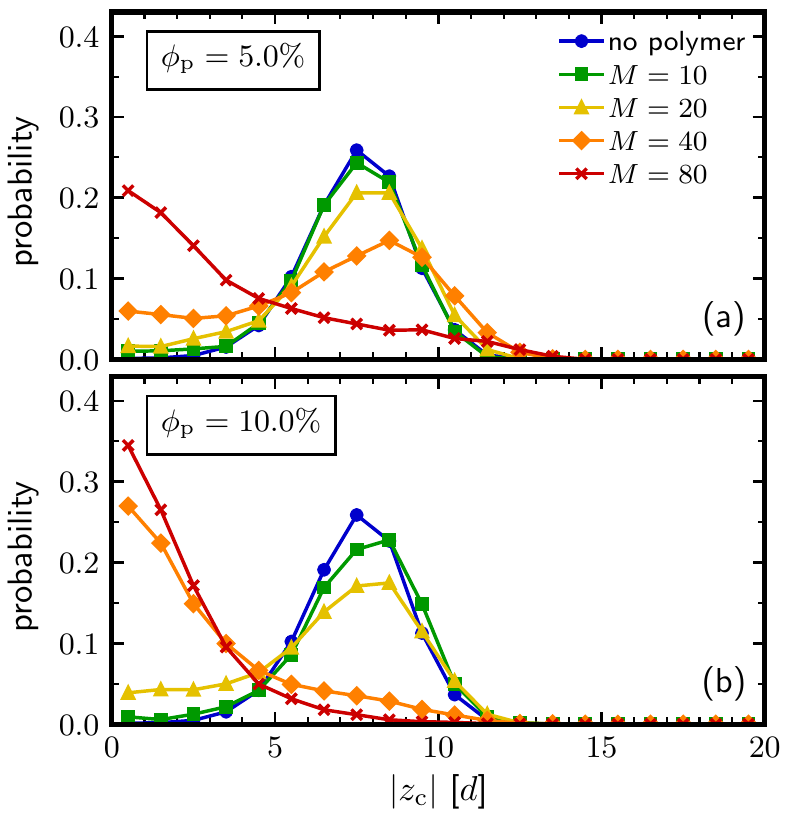}
    \caption{Distribution of droplet center-of-mass, $|z_{\rm c}|$, in a solution of polymers of length $M$
             at (a) $\phi_{\rm p} = 5.0\%$ and (b) $\phi_{\rm p} = 10.0\%$ for $f_x = 0.005\,\varepsilon/d$.}
    \label{fgr:distM}
\end{figure}

We next considered the impact of chain length $M$ on droplet focusing at two polymer concentrations, $\phi_{\rm p} = 5.0\%$
and $\phi_{\rm p} = 10.0\%$. Longer chains are expected to have better droplet focusing for three reasons:
(1) the elastic force should scale with $M$ \cite{Howard:2015bl}, (2) longer chains deform at lower shear rates and so act more
non-Newtonian, and (3) the maximum velocity was found to be lower for longer chains, reducing the outward inertial lift. The measured distributions
of $|z_{\rm c}|$ (Fig.~\ref{fgr:distM}) are clearly consistent with this hypothesis. The addition of polymers with $M=10$
did not have a significant impact on the droplet distribution compared to the neat solvent. This may not be surprising
given that ${\rm Wi} = 1.3$ for the $M = 10$ polymers, and the solution is nearly Newtonian.
However, adding polymers of increasing length improved the focusing onto the centerline in a monotonic fashion
for a given concentration.

We note, however, that there are additional concentration effects that influenced when polymers of a given size became effective focusers.
This is most apparent for the $M=40$ chains. At $\phi_{\rm p} = 5.0\%$, the droplet had a broad distribution
of $|z_{\rm c}|$ and a most probable position of $8.5\,d$. However, at $\phi_{\rm p} = 10.0\%$, the droplet
was strongly focused onto the centerline. We speculate that this difference in behavior is due to an
increase in elastic force with concentration, which was sufficient to overcome the inertial lift at $\phi_{\rm p} = 10.0\%$
but too weak at $\phi_{\rm p} = 5.0\%$.

\begin{figure}[h]
    \centering
    \includegraphics{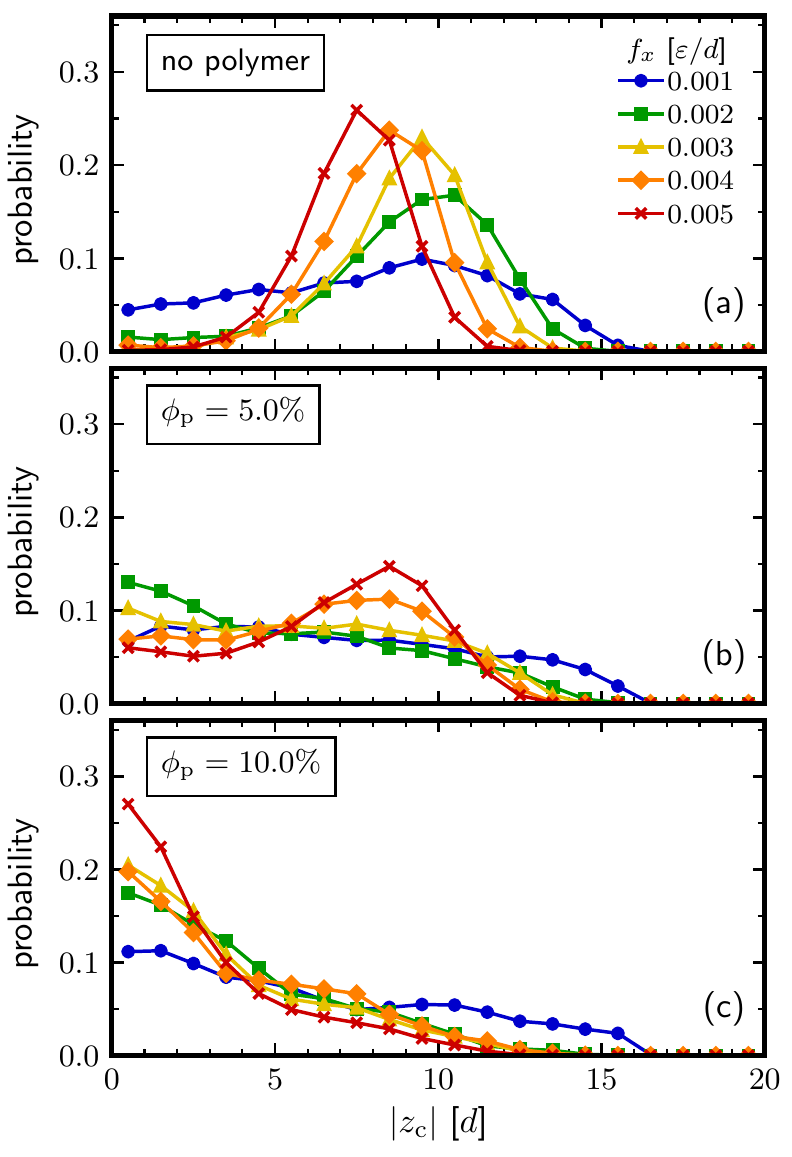}
    \caption{Distribution of droplet center-of-mass, $|z_{\rm c}|$, in a solution of $M=40$ polymers at
        (a) $\phi_{\rm p} = 5.0\%$ and (b) $\phi_{\rm p} = 10.0\%$ for $f_x = 0.005\,\varepsilon/d$.}
    \label{fgr:distfx}
\end{figure}

To understand this flow rate and concentration dependence in more detail, we computed the droplet distribution
for the $M=40$ chains at varying $f_x$ for $\phi_{\rm p} = 5.0\%$ and $10.0\%$, which we compare to the distributions
without any polymer (Fig.~\ref{fgr:distfx}). Without polymer, the droplet initially migrated outward as $f_x$ increased,
but the peak of this distribution moved inward with additional increases in $f_x$ (Fig.~\ref{fgr:distfx}a),
consistent with the simulations of Marson et al.\cite{Marson:2018fv}
At $\phi_{\rm p} = 5.0\%$ (Fig.~\ref{fgr:distfx}b), there was an initial trend to focus when $f_x \lesssim 0.003\,\varepsilon/d$. However,
at larger $f_x$, the droplet began to migrate outward, suggesting that inertial lift dominated over the available
elastic force. In contrast, the droplet distribution sharpened around the channel centerline at $\phi_{\rm p} = 10.0\%$
for all $f_x$ considered here (Fig.~\ref{fgr:distfx}c). It is possible that there is a sufficiently large $f_x$ that could exceed
the inward elastic force at this concentration. However, the droplet may
breakup under shear before such a force can be applied.

We finally tested the sensitivity of the viscoelastic focusing to the viscosity ratio between the droplet
and the solvent. In the Stokes flow limit, Chan and Leal showed that inward or outward droplet migration
can be obtained in a Newtonian solvent based on this ratio \cite{Chan:1979ep}. However, recent simulations by Marson et al.
suggest that such differences may not be as significant in the inertial regime \cite{Marson:2018fv}. Our primary interest
is in how the droplet focusing may change in a non-Newtonian polymer solution.
We varied the droplet viscosity ratio, $\mu_{\rm d} / \mu_{\rm s}$, from 0.54 ($\gamma_{\rm d} = 1.0\,m/\tau$)
to 5.3 ($\gamma_{\rm d} = 40.0\,m/\tau$) for the
$M=80$ polymer solutions at $\phi_{\rm p} = 10.0\%$, which focused the droplet when
$\mu_{\rm d}/\mu_{\rm s} = 1.0$. Since the effective viscosity of the polymer solution is higher
than that of the pure solvent, $\mu_{\rm d}/\mu_{\rm s}$ should be considered an upper bound on the viscosity ratio
between the droplet and the polymer solution.

We found no significant differences between the droplet distributions under
these conditions, and so we omit the data here for brevity.
This result may not be unexpected given the qualitative picture of the focusing mechanism.
The polymers are insoluble in the droplet, and so they primarily influence the fluid around it.
(The flow inside the droplet is affected by $\mu_{\rm d}$, but such effects may be secondary.)
Given that the viscosity ratio did not significantly alter the droplet distribution
in the inertial regime of the pure solvent for Marson et al. \cite{Marson:2018fv}, it is then not surprising that
the viscosity ratio also does not significantly change the droplet distribution in the polymer solution.
Indeed, letting $\mu_{\rm d}/\mu_{\rm s}\to\infty$ should recover the rigid particle limit of viscoelastic
focusing to which we have already drawn analogy. However, we do anticipate that the viscosity ratio may
still influence the droplet distribution more significantly in other flow regimes (e.g., Stokes flow limit)
that were not accessible to us in our simulations. In these cases, the migration forces controlled by the viscosity
ratio, with inward or outward direction \cite{Chan:1979ep}, would either work cooperatively or antagonistically
with the elastic force of a sufficiently deformed polymer.

\subsection{Droplet shape}
We have concentrated our discussion thus far on how polymers influence the droplet distribution in the microchannel,
but have not yet considered how the polymers influence the droplet shape and orientation in the flow.
To characterize the droplet shape, we first computed its gyration tensor $\mathbf{G}$,
\begin{equation}
G_{\alpha\beta} = \frac{1}{N_{\rm d}} \sum_{i=1}^{N_{\rm d}} \Delta r_{i,\alpha} \Delta r_{i,\beta},
\end{equation}
where $\Delta \mathbf{r}_i$ is the vector from the droplet center of mass to particle $i$, $\alpha$ and $\beta$
are indices in the usual tensor notation, and $N_{\rm d}$ is the number of particles in the droplet.
We then computed the eigenvalues $\lambda_\alpha^2$ of $\mathbf{G}$, whose corresponding eigenvectors give
the principle moments of the droplet, and sorted them in descending order, $\lambda_1^2 \ge \lambda_2^2 \ge \lambda_3^2$.
We determined the Taylor deformation parameter \cite{Taylor:1932tm,Taylor:1934wx}, a dimensionless measure of the asphericity of the droplet,
as
\begin{equation}
D \approx \frac{\lambda_1 - (\lambda_2+\lambda_3)/2}{\lambda_1+(\lambda_2+\lambda_3)/2}.
\end{equation}
For a sphere, $\mathbf{G}$ is diagonal ($\lambda_1 = \lambda_2 = \lambda_3$)
and $D = 0$, while a prolate spheroid has $\lambda_1 > \lambda_2 = \lambda_3$ and $D \to 1$ when the aspect
ratio between the major and minor axes of the spheroid increases. Hence, larger values of $D$ correspond to droplets that
have more significant deformation.
We additionally determined the inclination angle of the droplet relative to the flow direction, $\theta$, using $\mathbf{G}$ \cite{Ripoll:2006dh}:
\begin{equation}
\tan(2\theta) = \frac{2 G_{xz}}{G_{xx}-G_{zz}}.
\end{equation}
$\theta \approx 0^\circ$ for a sphere (no preferred orientation) or for an object completely aligned with
the flow, but $\theta \ne 0^\circ$ for particles that align with a relative tilt.

\begin{figure}[h]
    \centering
    \includegraphics{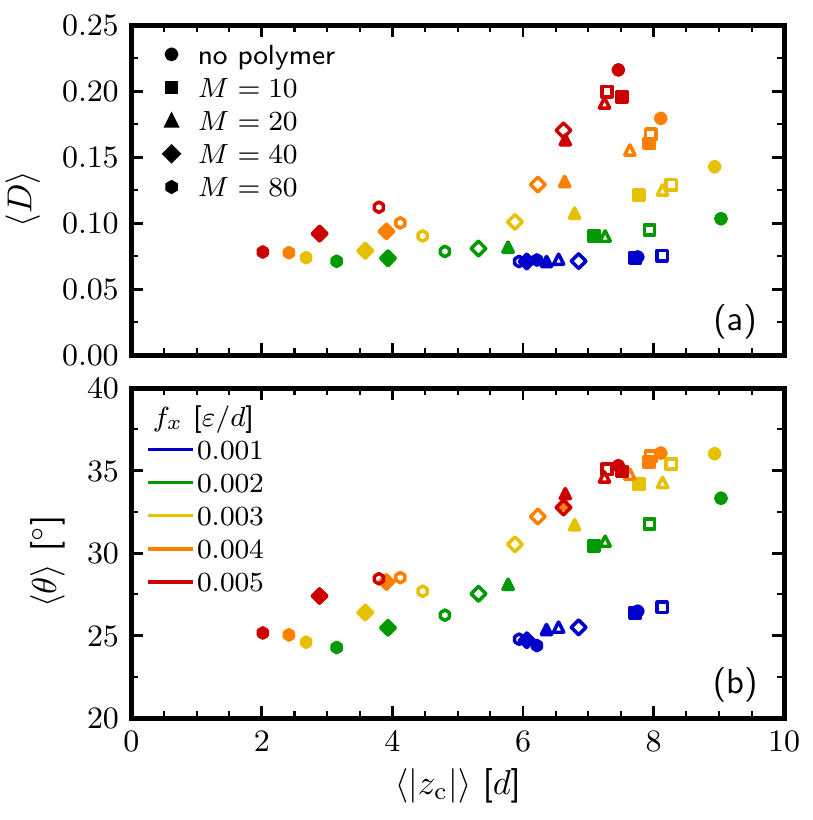}
    \caption{Average droplet (a) Taylor deformation parameter $\langle D \rangle$ and (b) inclination angle $\langle \theta \rangle$ versus
        average center-of-mass position $\langle |z_{\rm c}| \rangle$ for varied polymer chain lengths $M$ at $\phi_{\rm p}=5.0\%$ (open symbols)
        and 10.0\% (filled symbols). The solid lines are a guide to the eye for each applied body force $f_x$.
        }
    \label{fgr:deformz}
\end{figure}

It is well-established that $D$ increases with $\dot{\gamma}$ for a droplet in an unbounded shear flow \cite{Taylor:1932tm,Taylor:1934wx,Rallison:1978dd}.
In Poiseuille flows, the shear rate varies across the channel, and accordingly, the droplet may experience a different deformation
based on its lateral position.
Fig.~\ref{fgr:deformz}a shows the average deformation $\langle D \rangle$
versus the average center-of-mass position of the droplet $\langle|z_{\rm c}|\rangle$, clearly indicating that the droplet is (on average) more deformed
when it is (on average) farther from the centerline, where the shear rate is higher. The droplet is additionally
(on average) more inclined relative to the flow when it is more deformed (Fig.~\ref{fgr:deformz}b).

It is tempting to find a parameter to collapse the data in Fig.~\ref{fgr:deformz} onto a single curve,
e.g., using the capillary number \cite{Rallison:1978dd,Pan:2014bp,Marson:2018fv}. Unfortunately, such an analysis is again considerably complicated using
only average quantities because of the droplet distribution. Indeed, the average properties computed in
Fig.~\ref{fgr:deformz} and in prior studies\cite{Marson:2018fv} are intimately connected to the droplet distribution, which
sets the preferred droplet location and as a consequence, the shear rates it experiences.

\begin{figure}[h]
    \centering
    \includegraphics{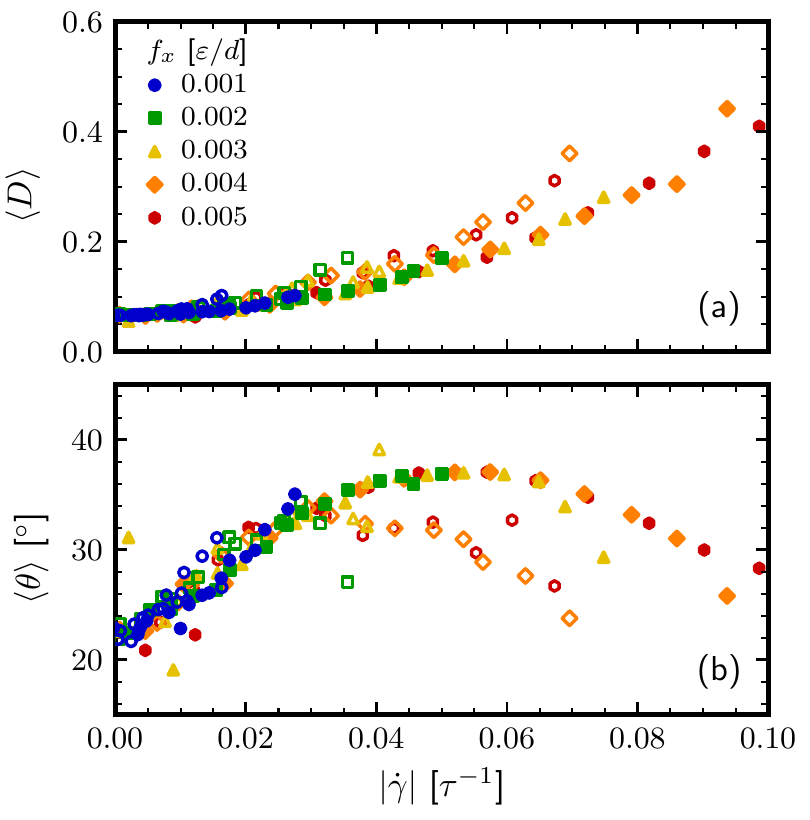}
    \caption{Average droplet (a) Taylor deformation parameter $\langle D \rangle$ and (b) inclination angle $\langle \theta \rangle$ versus
             shear rate at the droplet center-of-mass $|\dot{\gamma}|$ for neat solvent (closed symbols)
             and $M=80$ polymers with $\phi_{\rm p} = 10.0\%$ (open symbols). Note the collapse of the data
             for small shear rates (near channel center) with deviations at larger shear rates (near the walls).}
    \label{fgr:deformg}
\end{figure}

To deconvolve the droplet shape from the droplet distribution, we averaged $D$ and $\theta$ as functions
of the shear rate $|\dot{\gamma}|$ at the droplet center-of-mass using the flow fields measured in the simulations (Fig.~\ref{fgr:velocity}).
Figure \ref{fgr:deformg} shows the results of this procedure for the pure solvent and the $M=80$ polymers
with $\phi_{\rm p} = 10.0\%$. For a given polymer solution, the data were well-collapsed across all body
forces when plotted against $|\dot{\gamma}|$. Moreover, all data nearly collapsed onto a single curve
for small $\dot{\gamma}$, i.e., near the channel centerline. However, there
were some noticeable differences between different polymer solutions at larger $|\dot{\gamma}|$,
which corresponded to larger $f_x$ and positions closer to the channel walls. This discrepancy is not surprising
since in that regime wall effects on the droplet may be significant.

In general, the droplet deformation $D$ increased monotonically with shear rate (Fig.~\ref{fgr:deformg}a),
but never reached zero even as $\dot{\gamma} \to 0$ due to the finite size of the droplet. In contrast,
the inclination angle $\theta$ showed a maximum at intermediate shear rates (Fig.~\ref{fgr:deformg}b). On visual inspection of the
trajectories, this change in orientation with the flow appeared to be due to alignment of the droplet
with the walls. Other than these boundary effects, we found that the droplet
shape was not strongly influenced by the presence of the polymers in solution. Instead,
the deformation and orientation correlated strongly with the shear rate due to the imposed flow.
The primary roles of the polymers in setting the average deformation (Fig.~\ref{fgr:deformz}) were then
as viscosity modifiers, altering the flow field (Fig.~\ref{fgr:velocity}) for a given body force,
and as focusers for the droplet, which caused the droplet to experience a given shear rate with
higher probability.

\section{Conclusions}
We used dissipative particle dynamics computer simulations to show that a Brownian droplet in a dilute polymer solution
migrates toward the center of a parallel-plate microchannel under gravity-driven flow. The droplet had a distribution of
positions in the channel that sharpened near the center for longer polymer chains at higher concentrations, but had
a nontrivial dependence on the flow rate due to droplet and polymer deformation. The average droplet shape depended
on the droplet distribution because its local deformation was controlled by the shear rate. Our simulations demonstrate the
applicability of the viscoelastic focusing mechanism for Brownian droplets that are comparable in size to the polymer chains
in the viscoelastic fluid.

Polymer-induced migration may play an important role in droplet migration and mobility in small channels flooded with complex
fluids, such as those encountered in oil recovery from geological formations or in membrane filtration. In this work, we have neglected
polymer solubility and adsorption with the droplet, the presence of surfactants, complex microchannel boundaries, and collective interactions
between droplets at finite concentration. Such effects are surely present in many applications, and an intriguing
avenue of future research is to determine how they may assist or hinder polymer-induced droplet migration in microchannels.
Viscoelastic focusing will likely also influence the migration of other rigid and deformable objects in these complex fluid
mixtures, including colloids, star polymers \cite{Srivastva:2018ft}, dendrimers, cells, and micelles. Controlling the
distribution of these objects in a mixture through general inertial and viscoelastic focusing mechanisms presents an
opportunity to effect a separation.

\section*{Conflicts of interest}
There are no conflicts to declare.

\begin{acknowledgments}
We happily thank Florian M\"{u}ller-Plathe and Athanassios Panagiotopoulos for discussions inspiring this research.
Work by M.P.H. was supported as part of the Center for Materials for Water and Energy Systems, an Energy Frontier
Research Center funded by the U.S. Department of Energy, Office of Science, Basic Energy Sciences under Award
No. DE-SC0019272. T.M.T. acknowledges financial support from the Welch Foundation (Grant No. F-1696), and
A.N. acknowledges financial support from the German Research Foundation (DFG) under Project No. NI 1487-2/1.
The simulations are part of the Blue Waters sustained-petascale computing project, which
is supported by the National Science Foundation (awards OCI-0725070 and ACI-1238993)
and the state of Illinois. Blue Waters is a joint effort of the University of Illinois
at Urbana-Champaign and its National Center for Supercomputing Applications.
\end{acknowledgments}

\bibliography{droplet_focusing}

\begin{thebibliography}{85}%
\makeatletter
\providecommand \@ifxundefined [1]{%
 \@ifx{#1\undefined}
}%
\providecommand \@ifnum [1]{%
 \ifnum #1\expandafter \@firstoftwo
 \else \expandafter \@secondoftwo
 \fi
}%
\providecommand \@ifx [1]{%
 \ifx #1\expandafter \@firstoftwo
 \else \expandafter \@secondoftwo
 \fi
}%
\providecommand \natexlab [1]{#1}%
\providecommand \enquote  [1]{``#1''}%
\providecommand \bibnamefont  [1]{#1}%
\providecommand \bibfnamefont [1]{#1}%
\providecommand \citenamefont [1]{#1}%
\providecommand \href@noop [0]{\@secondoftwo}%
\providecommand \href [0]{\begingroup \@sanitize@url \@href}%
\providecommand \@href[1]{\@@startlink{#1}\@@href}%
\providecommand \@@href[1]{\endgroup#1\@@endlink}%
\providecommand \@sanitize@url [0]{\catcode `\\12\catcode `\$12\catcode
  `\&12\catcode `\#12\catcode `\^12\catcode `\_12\catcode `\%12\relax}%
\providecommand \@@startlink[1]{}%
\providecommand \@@endlink[0]{}%
\providecommand \url  [0]{\begingroup\@sanitize@url \@url }%
\providecommand \@url [1]{\endgroup\@href {#1}{\urlprefix }}%
\providecommand \urlprefix  [0]{URL }%
\providecommand \Eprint [0]{\href }%
\providecommand \doibase [0]{http://dx.doi.org/}%
\providecommand \selectlanguage [0]{\@gobble}%
\providecommand \bibinfo  [0]{\@secondoftwo}%
\providecommand \bibfield  [0]{\@secondoftwo}%
\providecommand \translation [1]{[#1]}%
\providecommand \BibitemOpen [0]{}%
\providecommand \bibitemStop [0]{}%
\providecommand \bibitemNoStop [0]{.\EOS\space}%
\providecommand \EOS [0]{\spacefactor3000\relax}%
\providecommand \BibitemShut  [1]{\csname bibitem#1\endcsname}%
\let\auto@bib@innerbib\@empty
%</preamble>
\bibitem [{\citenamefont {Stone}\ and\ \citenamefont
  {Kim}(2001)}]{Stone:2001ws}%
  \BibitemOpen
  \bibfield  {author} {\bibinfo {author} {\bibfnamefont {H.~A.}\ \bibnamefont
  {Stone}}\ and\ \bibinfo {author} {\bibfnamefont {S.}~\bibnamefont {Kim}},\
  }\href@noop {} {\bibfield  {journal} {\bibinfo  {journal} {AIChE J.}\
  }\textbf {\bibinfo {volume} {47}},\ \bibinfo {pages} {1250} (\bibinfo {year}
  {2001})}\BibitemShut {NoStop}%
\bibitem [{\citenamefont {Stone}, \citenamefont {Stroock},\ and\ \citenamefont
  {Ajdari}(2004)}]{Stone:2004ww}%
  \BibitemOpen
  \bibfield  {author} {\bibinfo {author} {\bibfnamefont {H.~A.}\ \bibnamefont
  {Stone}}, \bibinfo {author} {\bibfnamefont {A.~D.}\ \bibnamefont {Stroock}},
  \ and\ \bibinfo {author} {\bibfnamefont {A.}~\bibnamefont {Ajdari}},\
  }\href@noop {} {\bibfield  {journal} {\bibinfo  {journal} {Annu. Rev. Fluid
  Mech.}\ }\textbf {\bibinfo {volume} {36}},\ \bibinfo {pages} {381} (\bibinfo
  {year} {2004})}\BibitemShut {NoStop}%
\bibitem [{\citenamefont {Bhagat}, \citenamefont {Kuntaegowdanahalli},\ and\
  \citenamefont {Papautsky}(2008)}]{bhagat:physfluids:08}%
  \BibitemOpen
  \bibfield  {author} {\bibinfo {author} {\bibfnamefont {A.~A.~S.}\
  \bibnamefont {Bhagat}}, \bibinfo {author} {\bibfnamefont {S.~S.}\
  \bibnamefont {Kuntaegowdanahalli}}, \ and\ \bibinfo {author} {\bibfnamefont
  {I.}~\bibnamefont {Papautsky}},\ }\href@noop {} {\bibfield  {journal}
  {\bibinfo  {journal} {Phys. Fluids}\ }\textbf {\bibinfo {volume} {20}},\
  \bibinfo {pages} {101702} (\bibinfo {year} {2008})}\BibitemShut {NoStop}%
\bibitem [{\citenamefont {Hur}\ \emph {et~al.}(2011)\citenamefont {Hur},
  \citenamefont {Henderson-MacLennan}, \citenamefont {McCabe},\ and\
  \citenamefont {Di~Carlo}}]{Hur:2011vz}%
  \BibitemOpen
  \bibfield  {author} {\bibinfo {author} {\bibfnamefont {S.~C.}\ \bibnamefont
  {Hur}}, \bibinfo {author} {\bibfnamefont {N.~K.}\ \bibnamefont
  {Henderson-MacLennan}}, \bibinfo {author} {\bibfnamefont {E.~R.~B.}\
  \bibnamefont {McCabe}}, \ and\ \bibinfo {author} {\bibfnamefont
  {D.}~\bibnamefont {Di~Carlo}},\ }\href@noop {} {\bibfield  {journal}
  {\bibinfo  {journal} {Lab Chip}\ }\textbf {\bibinfo {volume} {11}},\ \bibinfo
  {pages} {912} (\bibinfo {year} {2011})}\BibitemShut {NoStop}%
\bibitem [{\citenamefont {Giddings}(1993)}]{Giddings:1993tq}%
  \BibitemOpen
  \bibfield  {author} {\bibinfo {author} {\bibfnamefont {J.~C.}\ \bibnamefont
  {Giddings}},\ }\href@noop {} {\bibfield  {journal} {\bibinfo  {journal}
  {Science}\ }\textbf {\bibinfo {volume} {260}},\ \bibinfo {pages} {1456}
  (\bibinfo {year} {1993})}\BibitemShut {NoStop}%
\bibitem [{\citenamefont {Kumar}\ and\ \citenamefont
  {Graham}(2012{\natexlab{a}})}]{Kumar:2012ev}%
  \BibitemOpen
  \bibfield  {author} {\bibinfo {author} {\bibfnamefont {A.}~\bibnamefont
  {Kumar}}\ and\ \bibinfo {author} {\bibfnamefont {M.~D.}\ \bibnamefont
  {Graham}},\ }\href@noop {} {\bibfield  {journal} {\bibinfo  {journal} {Phys.
  Rev. Lett.}\ }\textbf {\bibinfo {volume} {109}},\ \bibinfo {pages} {108102}
  (\bibinfo {year} {2012}{\natexlab{a}})}\BibitemShut {NoStop}%
\bibitem [{\citenamefont {Kumar}\ and\ \citenamefont
  {Graham}(2012{\natexlab{b}})}]{Kumar:2012en}%
  \BibitemOpen
  \bibfield  {author} {\bibinfo {author} {\bibfnamefont {A.}~\bibnamefont
  {Kumar}}\ and\ \bibinfo {author} {\bibfnamefont {M.~D.}\ \bibnamefont
  {Graham}},\ }\href@noop {} {\bibfield  {journal} {\bibinfo  {journal} {Soft
  Matter}\ }\textbf {\bibinfo {volume} {8}},\ \bibinfo {pages} {10536}
  (\bibinfo {year} {2012}{\natexlab{b}})}\BibitemShut {NoStop}%
\bibitem [{\citenamefont {Green}\ and\ \citenamefont
  {Willhite}(1998)}]{green:1998}%
  \BibitemOpen
  \bibfield  {author} {\bibinfo {author} {\bibfnamefont {D.~W.}\ \bibnamefont
  {Green}}\ and\ \bibinfo {author} {\bibfnamefont {G.~P.}\ \bibnamefont
  {Willhite}},\ }\enquote {\bibinfo {title} {Enhanced oil recovery},}\ \
  (\bibinfo  {publisher} {Society of Petroleum Engineers},\ \bibinfo {year}
  {1998})\ pp.\ \bibinfo {pages} {239--300}\BibitemShut {NoStop}%
\bibitem [{\citenamefont {Wilson~Jr.}(1977)}]{wilson:1977}%
  \BibitemOpen
  \bibfield  {author} {\bibinfo {author} {\bibfnamefont {L.~A.}\ \bibnamefont
  {Wilson~Jr.}},\ }\enquote {\bibinfo {title} {Improved oil recovery by
  surfactant and polymer flooding},}\ \ (\bibinfo  {publisher} {Academic
  Press},\ \bibinfo {year} {1977})\ pp.\ \bibinfo {pages} {1--26}\BibitemShut
  {NoStop}%
\bibitem [{\citenamefont {Tehrani}(1996)}]{tehrani:jrheol:96}%
  \BibitemOpen
  \bibfield  {author} {\bibinfo {author} {\bibfnamefont {M.~A.}\ \bibnamefont
  {Tehrani}},\ }\href@noop {} {\bibfield  {journal} {\bibinfo  {journal} {J.
  Rheol.}\ }\textbf {\bibinfo {volume} {40}},\ \bibinfo {pages} {1057}
  (\bibinfo {year} {1996})}\BibitemShut {NoStop}%
\bibitem [{\citenamefont {Fu}\ \emph {et~al.}(1999)\citenamefont {Fu},
  \citenamefont {Spence}, \citenamefont {Scherer}, \citenamefont {Arnold},\
  and\ \citenamefont {Quake}}]{fu:natbiotech:99}%
  \BibitemOpen
  \bibfield  {author} {\bibinfo {author} {\bibfnamefont {A.~Y.}\ \bibnamefont
  {Fu}}, \bibinfo {author} {\bibfnamefont {C.}~\bibnamefont {Spence}}, \bibinfo
  {author} {\bibfnamefont {A.}~\bibnamefont {Scherer}}, \bibinfo {author}
  {\bibfnamefont {F.~H.}\ \bibnamefont {Arnold}}, \ and\ \bibinfo {author}
  {\bibfnamefont {S.~R.}\ \bibnamefont {Quake}},\ }\href@noop {} {\bibfield
  {journal} {\bibinfo  {journal} {Nat. Biotech.}\ }\textbf {\bibinfo {volume}
  {17}},\ \bibinfo {pages} {1109} (\bibinfo {year} {1999})}\BibitemShut
  {NoStop}%
\bibitem [{\citenamefont {Cohen}(2005)}]{cohen:prl:05}%
  \BibitemOpen
  \bibfield  {author} {\bibinfo {author} {\bibfnamefont {A.~E.}\ \bibnamefont
  {Cohen}},\ }\href@noop {} {\bibfield  {journal} {\bibinfo  {journal} {Phys.
  Rev. Lett.}\ }\textbf {\bibinfo {volume} {94}},\ \bibinfo {pages} {118102}
  (\bibinfo {year} {2005})}\BibitemShut {NoStop}%
\bibitem [{\citenamefont {Lee}, \citenamefont {Purdom},\ and\ \citenamefont
  {Westervelt}(2004)}]{lee:appl:04}%
  \BibitemOpen
  \bibfield  {author} {\bibinfo {author} {\bibfnamefont {H.}~\bibnamefont
  {Lee}}, \bibinfo {author} {\bibfnamefont {A.~M.}\ \bibnamefont {Purdom}}, \
  and\ \bibinfo {author} {\bibfnamefont {R.~M.}\ \bibnamefont {Westervelt}},\
  }\href@noop {} {\bibfield  {journal} {\bibinfo  {journal} {Appl. Phys.
  Lett.}\ }\textbf {\bibinfo {volume} {85}},\ \bibinfo {pages} {1063} (\bibinfo
  {year} {2004})}\BibitemShut {NoStop}%
\bibitem [{\citenamefont {Das}, \citenamefont {Mandal},\ and\ \citenamefont
  {Chakraborty}(2018)}]{Das:2017eq}%
  \BibitemOpen
  \bibfield  {author} {\bibinfo {author} {\bibfnamefont {S.}~\bibnamefont
  {Das}}, \bibinfo {author} {\bibfnamefont {S.}~\bibnamefont {Mandal}}, \ and\
  \bibinfo {author} {\bibfnamefont {S.}~\bibnamefont {Chakraborty}},\
  }\href@noop {} {\bibfield  {journal} {\bibinfo  {journal} {J. Fluid Mech.}\
  }\textbf {\bibinfo {volume} {835}},\ \bibinfo {pages} {170} (\bibinfo {year}
  {2018})}\BibitemShut {NoStop}%
\bibitem [{\citenamefont {Das}\ and\ \citenamefont
  {Chakraborty}(2018)}]{Das:2018dn}%
  \BibitemOpen
  \bibfield  {author} {\bibinfo {author} {\bibfnamefont {S.}~\bibnamefont
  {Das}}\ and\ \bibinfo {author} {\bibfnamefont {S.}~\bibnamefont
  {Chakraborty}},\ }\href@noop {} {\bibfield  {journal} {\bibinfo  {journal}
  {Phys. Rev. Fluids}\ }\textbf {\bibinfo {volume} {3}},\ \bibinfo {pages}
  {103602} (\bibinfo {year} {2018})}\BibitemShut {NoStop}%
\bibitem [{\citenamefont {Karimi}, \citenamefont {Yazdi},\ and\ \citenamefont
  {Ardekani}(2013)}]{karimi:biofluidics:13}%
  \BibitemOpen
  \bibfield  {author} {\bibinfo {author} {\bibfnamefont {A.}~\bibnamefont
  {Karimi}}, \bibinfo {author} {\bibfnamefont {S.}~\bibnamefont {Yazdi}}, \
  and\ \bibinfo {author} {\bibfnamefont {A.~M.}\ \bibnamefont {Ardekani}},\
  }\href@noop {} {\bibfield  {journal} {\bibinfo  {journal} {Biomicrofluidics}\
  }\textbf {\bibinfo {volume} {7}},\ \bibinfo {pages} {021501} (\bibinfo {year}
  {2013})}\BibitemShut {NoStop}%
\bibitem [{\citenamefont {Amini}, \citenamefont {Lee},\ and\ \citenamefont
  {Di~Carlo}(2014)}]{amini:labchip:14}%
  \BibitemOpen
  \bibfield  {author} {\bibinfo {author} {\bibfnamefont {H.}~\bibnamefont
  {Amini}}, \bibinfo {author} {\bibfnamefont {W.}~\bibnamefont {Lee}}, \ and\
  \bibinfo {author} {\bibfnamefont {D.}~\bibnamefont {Di~Carlo}},\ }\href@noop
  {} {\bibfield  {journal} {\bibinfo  {journal} {Lab Chip}\ }\textbf {\bibinfo
  {volume} {14}},\ \bibinfo {pages} {2739} (\bibinfo {year}
  {2014})}\BibitemShut {NoStop}%
\bibitem [{\citenamefont {Leal}(1980)}]{Leal:1980wy}%
  \BibitemOpen
  \bibfield  {author} {\bibinfo {author} {\bibfnamefont {L.~G.}\ \bibnamefont
  {Leal}},\ }\href@noop {} {\bibfield  {journal} {\bibinfo  {journal} {Ann.
  Rev. Fluid Mech.}\ }\textbf {\bibinfo {volume} {12}},\ \bibinfo {pages} {435}
  (\bibinfo {year} {1980})}\BibitemShut {NoStop}%
\bibitem [{\citenamefont {Stone}(2000)}]{Stone:2000js}%
  \BibitemOpen
  \bibfield  {author} {\bibinfo {author} {\bibfnamefont {H.~A.}\ \bibnamefont
  {Stone}},\ }\href@noop {} {\bibfield  {journal} {\bibinfo  {journal} {J.
  Fluid Mech.}\ }\textbf {\bibinfo {volume} {409}},\ \bibinfo {pages} {165}
  (\bibinfo {year} {2000})}\BibitemShut {NoStop}%
\bibitem [{\citenamefont {Matas}, \citenamefont {Morris},\ and\ \citenamefont
  {Guazzelli}(2004)}]{matas:jfluid:04}%
  \BibitemOpen
  \bibfield  {author} {\bibinfo {author} {\bibfnamefont {J.-P.}\ \bibnamefont
  {Matas}}, \bibinfo {author} {\bibfnamefont {J.~F.}\ \bibnamefont {Morris}}, \
  and\ \bibinfo {author} {\bibfnamefont {{\' E}.}~\bibnamefont {Guazzelli}},\
  }\href@noop {} {\bibfield  {journal} {\bibinfo  {journal} {J. Fluid. Mech.}\
  }\textbf {\bibinfo {volume} {515}},\ \bibinfo {pages} {171} (\bibinfo {year}
  {2004})}\BibitemShut {NoStop}%
\bibitem [{\citenamefont {Di~Carlo}\ \emph {et~al.}(2007)\citenamefont
  {Di~Carlo}, \citenamefont {Irimia}, \citenamefont {Tompkins},\ and\
  \citenamefont {Toner}}]{dicarlo:pnas:07}%
  \BibitemOpen
  \bibfield  {author} {\bibinfo {author} {\bibfnamefont {D.}~\bibnamefont
  {Di~Carlo}}, \bibinfo {author} {\bibfnamefont {D.}~\bibnamefont {Irimia}},
  \bibinfo {author} {\bibfnamefont {R.~G.}\ \bibnamefont {Tompkins}}, \ and\
  \bibinfo {author} {\bibfnamefont {M.}~\bibnamefont {Toner}},\ }\href@noop {}
  {\bibfield  {journal} {\bibinfo  {journal} {Proc. Natl. Acad. Sci. USA}\
  }\textbf {\bibinfo {volume} {104}},\ \bibinfo {pages} {18892} (\bibinfo
  {year} {2007})}\BibitemShut {NoStop}%
\bibitem [{\citenamefont {Segr\'{e}}\ and\ \citenamefont
  {Silberberg}(1961)}]{segre:nature:61}%
  \BibitemOpen
  \bibfield  {author} {\bibinfo {author} {\bibfnamefont {G.}~\bibnamefont
  {Segr\'{e}}}\ and\ \bibinfo {author} {\bibfnamefont {A.}~\bibnamefont
  {Silberberg}},\ }\href@noop {} {\bibfield  {journal} {\bibinfo  {journal}
  {Nature}\ }\textbf {\bibinfo {volume} {189}},\ \bibinfo {pages} {209}
  (\bibinfo {year} {1961})}\BibitemShut {NoStop}%
\bibitem [{\citenamefont {Ho}\ and\ \citenamefont {~}(1974)}]{Ho:1974gm}%
  \BibitemOpen
  \bibfield  {author} {\bibinfo {author} {\bibfnamefont {B.~P.}\ \bibnamefont
  {Ho}}\ and\ \bibinfo {author} {\bibfnamefont {L.~G.}\ \bibnamefont {~}},\
  }\href@noop {} {\bibfield  {journal} {\bibinfo  {journal} {J. Fluid Mech.}\
  }\textbf {\bibinfo {volume} {65}},\ \bibinfo {pages} {365} (\bibinfo {year}
  {1974})}\BibitemShut {NoStop}%
\bibitem [{\citenamefont {Di~Carlo}\ \emph {et~al.}(2009)\citenamefont
  {Di~Carlo}, \citenamefont {Edd}, \citenamefont {Humphry}, \citenamefont
  {Stone},\ and\ \citenamefont {Toner}}]{dicarlo:prl:2009}%
  \BibitemOpen
  \bibfield  {author} {\bibinfo {author} {\bibfnamefont {D.}~\bibnamefont
  {Di~Carlo}}, \bibinfo {author} {\bibfnamefont {J.~F.}\ \bibnamefont {Edd}},
  \bibinfo {author} {\bibfnamefont {K.~J.}\ \bibnamefont {Humphry}}, \bibinfo
  {author} {\bibfnamefont {H.~A.}\ \bibnamefont {Stone}}, \ and\ \bibinfo
  {author} {\bibfnamefont {M.}~\bibnamefont {Toner}},\ }\href@noop {}
  {\bibfield  {journal} {\bibinfo  {journal} {Phys. Rev. Lett.}\ }\textbf
  {\bibinfo {volume} {102}},\ \bibinfo {pages} {094503} (\bibinfo {year}
  {2009})}\BibitemShut {NoStop}%
\bibitem [{\citenamefont {Chan}\ and\ \citenamefont
  {Leal}(1979)}]{Chan:1979ep}%
  \BibitemOpen
  \bibfield  {author} {\bibinfo {author} {\bibfnamefont {P.~C.-H.}\
  \bibnamefont {Chan}}\ and\ \bibinfo {author} {\bibfnamefont {L.~G.}\
  \bibnamefont {Leal}},\ }\href@noop {} {\bibfield  {journal} {\bibinfo
  {journal} {J. Fluid Mech.}\ }\textbf {\bibinfo {volume} {92}},\ \bibinfo
  {pages} {131} (\bibinfo {year} {1979})}\BibitemShut {NoStop}%
\bibitem [{\citenamefont {Stan}\ \emph {et~al.}(2011)\citenamefont {Stan},
  \citenamefont {Guglielmini}, \citenamefont {Ellerbee}, \citenamefont
  {Caviezel}, \citenamefont {Stone},\ and\ \citenamefont
  {Whitesides}}]{Stan:2011ez}%
  \BibitemOpen
  \bibfield  {author} {\bibinfo {author} {\bibfnamefont {C.~A.}\ \bibnamefont
  {Stan}}, \bibinfo {author} {\bibfnamefont {L.}~\bibnamefont {Guglielmini}},
  \bibinfo {author} {\bibfnamefont {A.~K.}\ \bibnamefont {Ellerbee}}, \bibinfo
  {author} {\bibfnamefont {D.}~\bibnamefont {Caviezel}}, \bibinfo {author}
  {\bibfnamefont {H.~A.}\ \bibnamefont {Stone}}, \ and\ \bibinfo {author}
  {\bibfnamefont {G.~M.}\ \bibnamefont {Whitesides}},\ }\href@noop {}
  {\bibfield  {journal} {\bibinfo  {journal} {Phys. Rev. E}\ }\textbf {\bibinfo
  {volume} {84}},\ \bibinfo {pages} {036302} (\bibinfo {year}
  {2011})}\BibitemShut {NoStop}%
\bibitem [{\citenamefont {Stan}\ \emph {et~al.}(2013)\citenamefont {Stan},
  \citenamefont {Ellerbee}, \citenamefont {Guglielmini}, \citenamefont
  {Stone},\ and\ \citenamefont {Whitesides}}]{Stan:2013cv}%
  \BibitemOpen
  \bibfield  {author} {\bibinfo {author} {\bibfnamefont {C.~A.}\ \bibnamefont
  {Stan}}, \bibinfo {author} {\bibfnamefont {A.~K.}\ \bibnamefont {Ellerbee}},
  \bibinfo {author} {\bibfnamefont {L.}~\bibnamefont {Guglielmini}}, \bibinfo
  {author} {\bibfnamefont {H.~A.}\ \bibnamefont {Stone}}, \ and\ \bibinfo
  {author} {\bibfnamefont {G.~M.}\ \bibnamefont {Whitesides}},\ }\href@noop {}
  {\bibfield  {journal} {\bibinfo  {journal} {Lab Chip}\ }\textbf {\bibinfo
  {volume} {13}},\ \bibinfo {pages} {365} (\bibinfo {year} {2013})}\BibitemShut
  {NoStop}%
\bibitem [{\citenamefont {Legendre}\ and\ \citenamefont
  {Magnaudet}(1997)}]{Legendre:1997bh}%
  \BibitemOpen
  \bibfield  {author} {\bibinfo {author} {\bibfnamefont {D.}~\bibnamefont
  {Legendre}}\ and\ \bibinfo {author} {\bibfnamefont {J.}~\bibnamefont
  {Magnaudet}},\ }\href@noop {} {\bibfield  {journal} {\bibinfo  {journal}
  {Phys. Fluids}\ }\textbf {\bibinfo {volume} {9}},\ \bibinfo {pages} {3572}
  (\bibinfo {year} {1997})}\BibitemShut {NoStop}%
\bibitem [{\citenamefont {Saffman}(1965)}]{Saffman:1965di}%
  \BibitemOpen
  \bibfield  {author} {\bibinfo {author} {\bibfnamefont {P.~G.}\ \bibnamefont
  {Saffman}},\ }\href@noop {} {\bibfield  {journal} {\bibinfo  {journal} {J.
  Fluid Mech.}\ }\textbf {\bibinfo {volume} {22}},\ \bibinfo {pages} {385}
  (\bibinfo {year} {1965})}\BibitemShut {NoStop}%
\bibitem [{\citenamefont {Saffman}(1968)}]{Saffman:1968jk}%
  \BibitemOpen
  \bibfield  {author} {\bibinfo {author} {\bibfnamefont {P.~G.}\ \bibnamefont
  {Saffman}},\ }\href@noop {} {\bibfield  {journal} {\bibinfo  {journal} {J.
  Fluid Mech.}\ }\textbf {\bibinfo {volume} {31}},\ \bibinfo {pages} {624}
  (\bibinfo {year} {1968})}\BibitemShut {NoStop}%
\bibitem [{\citenamefont {Karnis}, \citenamefont {Goldsmith},\ and\
  \citenamefont {Mason}(1966)}]{Karnis:1966va}%
  \BibitemOpen
  \bibfield  {author} {\bibinfo {author} {\bibfnamefont {A.}~\bibnamefont
  {Karnis}}, \bibinfo {author} {\bibfnamefont {H.~L.}\ \bibnamefont
  {Goldsmith}}, \ and\ \bibinfo {author} {\bibfnamefont {S.~G.}\ \bibnamefont
  {Mason}},\ }\href@noop {} {\bibfield  {journal} {\bibinfo  {journal} {Can. J.
  Chem. Eng.}\ }\textbf {\bibinfo {volume} {44}},\ \bibinfo {pages} {181}
  (\bibinfo {year} {1966})}\BibitemShut {NoStop}%
\bibitem [{\citenamefont {Mortazavi}\ and\ \citenamefont
  {Tryggvason}(2000)}]{Mortazavi:2000cv}%
  \BibitemOpen
  \bibfield  {author} {\bibinfo {author} {\bibfnamefont {S.}~\bibnamefont
  {Mortazavi}}\ and\ \bibinfo {author} {\bibfnamefont {G.}~\bibnamefont
  {Tryggvason}},\ }\href@noop {} {\bibfield  {journal} {\bibinfo  {journal} {J.
  Fluid Mech.}\ }\textbf {\bibinfo {volume} {411}},\ \bibinfo {pages} {325}
  (\bibinfo {year} {2000})}\BibitemShut {NoStop}%
\bibitem [{\citenamefont {Chen}\ \emph {et~al.}(2014)\citenamefont {Chen},
  \citenamefont {Xue}, \citenamefont {Zhang}, \citenamefont {Hu}, \citenamefont
  {Jiang},\ and\ \citenamefont {Sun}}]{Chen:2014ve}%
  \BibitemOpen
  \bibfield  {author} {\bibinfo {author} {\bibfnamefont {X.}~\bibnamefont
  {Chen}}, \bibinfo {author} {\bibfnamefont {C.}~\bibnamefont {Xue}}, \bibinfo
  {author} {\bibfnamefont {L.}~\bibnamefont {Zhang}}, \bibinfo {author}
  {\bibfnamefont {G.}~\bibnamefont {Hu}}, \bibinfo {author} {\bibfnamefont
  {X.}~\bibnamefont {Jiang}}, \ and\ \bibinfo {author} {\bibfnamefont
  {J.}~\bibnamefont {Sun}},\ }\href@noop {} {\bibfield  {journal} {\bibinfo
  {journal} {Phys. Fluids}\ }\textbf {\bibinfo {volume} {26}},\ \bibinfo
  {pages} {112003} (\bibinfo {year} {2014})}\BibitemShut {NoStop}%
\bibitem [{\citenamefont {Pan}\ \emph {et~al.}(2016)\citenamefont {Pan},
  \citenamefont {Lin}, \citenamefont {Zhang},\ and\ \citenamefont
  {Shao}}]{Pan:2016cd}%
  \BibitemOpen
  \bibfield  {author} {\bibinfo {author} {\bibfnamefont {D.-y.}\ \bibnamefont
  {Pan}}, \bibinfo {author} {\bibfnamefont {Y.-q.}\ \bibnamefont {Lin}},
  \bibinfo {author} {\bibfnamefont {L.-x.}\ \bibnamefont {Zhang}}, \ and\
  \bibinfo {author} {\bibfnamefont {X.-m.}\ \bibnamefont {Shao}},\ }\href@noop
  {} {\bibfield  {journal} {\bibinfo  {journal} {J. Hydrodyn., Ser. B}\
  }\textbf {\bibinfo {volume} {28}},\ \bibinfo {pages} {702} (\bibinfo {year}
  {2016})}\BibitemShut {NoStop}%
\bibitem [{\citenamefont {Marson}\ \emph {et~al.}(2018)\citenamefont {Marson},
  \citenamefont {Huang}, \citenamefont {Huang}, \citenamefont {Fu},\ and\
  \citenamefont {Larson}}]{Marson:2018fv}%
  \BibitemOpen
  \bibfield  {author} {\bibinfo {author} {\bibfnamefont {R.~L.}\ \bibnamefont
  {Marson}}, \bibinfo {author} {\bibfnamefont {Y.}~\bibnamefont {Huang}},
  \bibinfo {author} {\bibfnamefont {M.}~\bibnamefont {Huang}}, \bibinfo
  {author} {\bibfnamefont {T.}~\bibnamefont {Fu}}, \ and\ \bibinfo {author}
  {\bibfnamefont {R.~G.}\ \bibnamefont {Larson}},\ }\href@noop {} {\bibfield
  {journal} {\bibinfo  {journal} {Soft Matter}\ }\textbf {\bibinfo {volume}
  {14}},\ \bibinfo {pages} {2267} (\bibinfo {year} {2018})}\BibitemShut
  {NoStop}%
\bibitem [{\citenamefont {D'Avino}, \citenamefont {Greco},\ and\ \citenamefont
  {Maffettone}(2017)}]{DAvino:2017kh}%
  \BibitemOpen
  \bibfield  {author} {\bibinfo {author} {\bibfnamefont {G.}~\bibnamefont
  {D'Avino}}, \bibinfo {author} {\bibfnamefont {F.}~\bibnamefont {Greco}}, \
  and\ \bibinfo {author} {\bibfnamefont {P.~L.}\ \bibnamefont {Maffettone}},\
  }\href@noop {} {\bibfield  {journal} {\bibinfo  {journal} {Ann. Rev. Fluid
  Mech.}\ }\textbf {\bibinfo {volume} {49}},\ \bibinfo {pages} {341} (\bibinfo
  {year} {2017})}\BibitemShut {NoStop}%
\bibitem [{\citenamefont {Leshansky}\ \emph {et~al.}(2007)\citenamefont
  {Leshansky}, \citenamefont {Bransky}, \citenamefont {Korin},\ and\
  \citenamefont {Dinnar}}]{leshansky:prl:07}%
  \BibitemOpen
  \bibfield  {author} {\bibinfo {author} {\bibfnamefont {A.~M.}\ \bibnamefont
  {Leshansky}}, \bibinfo {author} {\bibfnamefont {A.}~\bibnamefont {Bransky}},
  \bibinfo {author} {\bibfnamefont {N.}~\bibnamefont {Korin}}, \ and\ \bibinfo
  {author} {\bibfnamefont {U.}~\bibnamefont {Dinnar}},\ }\href@noop {}
  {\bibfield  {journal} {\bibinfo  {journal} {Phys. Rev. Lett.}\ }\textbf
  {\bibinfo {volume} {98}},\ \bibinfo {pages} {234501} (\bibinfo {year}
  {2007})}\BibitemShut {NoStop}%
\bibitem [{\citenamefont {Karnis}\ and\ \citenamefont
  {Mason}(1966)}]{Karnis:1966tp}%
  \BibitemOpen
  \bibfield  {author} {\bibinfo {author} {\bibfnamefont {A.}~\bibnamefont
  {Karnis}}\ and\ \bibinfo {author} {\bibfnamefont {S.~G.}\ \bibnamefont
  {Mason}},\ }\href@noop {} {\bibfield  {journal} {\bibinfo  {journal} {Trans.
  Soc. Rheo.}\ }\textbf {\bibinfo {volume} {10}},\ \bibinfo {pages} {571}
  (\bibinfo {year} {1966})}\BibitemShut {NoStop}%
\bibitem [{\citenamefont {Gauthier}, \citenamefont {Goldsmith},\ and\
  \citenamefont {Mason}(1971{\natexlab{a}})}]{Gauthier:1971wr}%
  \BibitemOpen
  \bibfield  {author} {\bibinfo {author} {\bibfnamefont {F.}~\bibnamefont
  {Gauthier}}, \bibinfo {author} {\bibfnamefont {H.~L.}\ \bibnamefont
  {Goldsmith}}, \ and\ \bibinfo {author} {\bibfnamefont {S.~G.}\ \bibnamefont
  {Mason}},\ }\href@noop {} {\bibfield  {journal} {\bibinfo  {journal} {Rheo.
  Acta}\ }\textbf {\bibinfo {volume} {10}},\ \bibinfo {pages} {344} (\bibinfo
  {year} {1971}{\natexlab{a}})}\BibitemShut {NoStop}%
\bibitem [{\citenamefont {Gauthier}, \citenamefont {Goldsmith},\ and\
  \citenamefont {Mason}(1971{\natexlab{b}})}]{Gauthier:1971vr}%
  \BibitemOpen
  \bibfield  {author} {\bibinfo {author} {\bibfnamefont {F.}~\bibnamefont
  {Gauthier}}, \bibinfo {author} {\bibfnamefont {H.~L.}\ \bibnamefont
  {Goldsmith}}, \ and\ \bibinfo {author} {\bibfnamefont {S.~G.}\ \bibnamefont
  {Mason}},\ }\href@noop {} {\bibfield  {journal} {\bibinfo  {journal} {Trans.
  Soc. Rheo.}\ }\textbf {\bibinfo {volume} {15}},\ \bibinfo {pages} {297}
  (\bibinfo {year} {1971}{\natexlab{b}})}\BibitemShut {NoStop}%
\bibitem [{\citenamefont {Kim}\ \emph {et~al.}(2012)\citenamefont {Kim},
  \citenamefont {Ahn}, \citenamefont {Lee},\ and\ \citenamefont
  {Kim}}]{kim:labchip:12}%
  \BibitemOpen
  \bibfield  {author} {\bibinfo {author} {\bibfnamefont {J.~Y.}\ \bibnamefont
  {Kim}}, \bibinfo {author} {\bibfnamefont {S.~W.}\ \bibnamefont {Ahn}},
  \bibinfo {author} {\bibfnamefont {S.~S.}\ \bibnamefont {Lee}}, \ and\
  \bibinfo {author} {\bibfnamefont {J.~M.}\ \bibnamefont {Kim}},\ }\href@noop
  {} {\bibfield  {journal} {\bibinfo  {journal} {Lab Chip}\ }\textbf {\bibinfo
  {volume} {12}},\ \bibinfo {pages} {2807} (\bibinfo {year}
  {2012})}\BibitemShut {NoStop}%
\bibitem [{\citenamefont {Kim}\ and\ \citenamefont {Kim}(2016)}]{Kim:2016tg}%
  \BibitemOpen
  \bibfield  {author} {\bibinfo {author} {\bibfnamefont {B.}~\bibnamefont
  {Kim}}\ and\ \bibinfo {author} {\bibfnamefont {J.~M.}\ \bibnamefont {Kim}},\
  }\href@noop {} {\bibfield  {journal} {\bibinfo  {journal} {Biomicrofluidics}\
  }\textbf {\bibinfo {volume} {10}},\ \bibinfo {pages} {024111} (\bibinfo
  {year} {2016})}\BibitemShut {NoStop}%
\bibitem [{\citenamefont {D'Avino}\ \emph {et~al.}(2012)\citenamefont
  {D'Avino}, \citenamefont {Romeo}, \citenamefont {Villone}, \citenamefont
  {Greco}, \citenamefont {Netti},\ and\ \citenamefont
  {Maffettone}}]{davino:labchip:2012}%
  \BibitemOpen
  \bibfield  {author} {\bibinfo {author} {\bibfnamefont {G.}~\bibnamefont
  {D'Avino}}, \bibinfo {author} {\bibfnamefont {G.}~\bibnamefont {Romeo}},
  \bibinfo {author} {\bibfnamefont {M.~M.}\ \bibnamefont {Villone}}, \bibinfo
  {author} {\bibfnamefont {F.}~\bibnamefont {Greco}}, \bibinfo {author}
  {\bibfnamefont {P.~A.}\ \bibnamefont {Netti}}, \ and\ \bibinfo {author}
  {\bibfnamefont {P.~L.}\ \bibnamefont {Maffettone}},\ }\href@noop {}
  {\bibfield  {journal} {\bibinfo  {journal} {Lab Chip}\ }\textbf {\bibinfo
  {volume} {12}},\ \bibinfo {pages} {1638} (\bibinfo {year}
  {2012})}\BibitemShut {NoStop}%
\bibitem [{\citenamefont {De~Santo}\ \emph {et~al.}(2014)\citenamefont
  {De~Santo}, \citenamefont {D'Avino}, \citenamefont {Romeo}, \citenamefont
  {Greco}, \citenamefont {Netti},\ and\ \citenamefont
  {Maffettone}}]{santo:prapp:2014}%
  \BibitemOpen
  \bibfield  {author} {\bibinfo {author} {\bibfnamefont {I.}~\bibnamefont
  {De~Santo}}, \bibinfo {author} {\bibfnamefont {G.}~\bibnamefont {D'Avino}},
  \bibinfo {author} {\bibfnamefont {G.}~\bibnamefont {Romeo}}, \bibinfo
  {author} {\bibfnamefont {F.}~\bibnamefont {Greco}}, \bibinfo {author}
  {\bibfnamefont {P.~A.}\ \bibnamefont {Netti}}, \ and\ \bibinfo {author}
  {\bibfnamefont {P.~L.}\ \bibnamefont {Maffettone}},\ }\href@noop {}
  {\bibfield  {journal} {\bibinfo  {journal} {Phys. Rev. Appl.}\ }\textbf
  {\bibinfo {volume} {2}},\ \bibinfo {pages} {064001} (\bibinfo {year}
  {2014})}\BibitemShut {NoStop}%
\bibitem [{\citenamefont {Nikoubashman}\ \emph
  {et~al.}(2014{\natexlab{a}})\citenamefont {Nikoubashman}, \citenamefont
  {Mahynski}, \citenamefont {Pirayandeh},\ and\ \citenamefont
  {Panagiotopoulos}}]{nikoubashman:jcp:14}%
  \BibitemOpen
  \bibfield  {author} {\bibinfo {author} {\bibfnamefont {A.}~\bibnamefont
  {Nikoubashman}}, \bibinfo {author} {\bibfnamefont {N.~A.}\ \bibnamefont
  {Mahynski}}, \bibinfo {author} {\bibfnamefont {A.~H.}\ \bibnamefont
  {Pirayandeh}}, \ and\ \bibinfo {author} {\bibfnamefont {A.~Z.}\ \bibnamefont
  {Panagiotopoulos}},\ }\href@noop {} {\bibfield  {journal} {\bibinfo
  {journal} {J. Chem. Phys.}\ }\textbf {\bibinfo {volume} {140}},\ \bibinfo
  {pages} {094903} (\bibinfo {year} {2014}{\natexlab{a}})}\BibitemShut
  {NoStop}%
\bibitem [{\citenamefont {Nikoubashman}\ \emph
  {et~al.}(2014{\natexlab{b}})\citenamefont {Nikoubashman}, \citenamefont
  {Mahynski}, \citenamefont {Howard},\ and\ \citenamefont
  {Panagiotopoulos}}]{nikoubashman:erratum:2014}%
  \BibitemOpen
  \bibfield  {author} {\bibinfo {author} {\bibfnamefont {A.}~\bibnamefont
  {Nikoubashman}}, \bibinfo {author} {\bibfnamefont {N.~A.}\ \bibnamefont
  {Mahynski}}, \bibinfo {author} {\bibfnamefont {M.~P.}\ \bibnamefont
  {Howard}}, \ and\ \bibinfo {author} {\bibfnamefont {A.~Z.}\ \bibnamefont
  {Panagiotopoulos}},\ }\href@noop {} {\bibfield  {journal} {\bibinfo
  {journal} {J. Chem. Phys.}\ }\textbf {\bibinfo {volume} {141}},\ \bibinfo
  {pages} {149906} (\bibinfo {year} {2014}{\natexlab{b}})}\BibitemShut
  {NoStop}%
\bibitem [{\citenamefont {Howard}, \citenamefont {Panagiotopoulos},\ and\
  \citenamefont {Nikoubashman}(2015)}]{Howard:2015bl}%
  \BibitemOpen
  \bibfield  {author} {\bibinfo {author} {\bibfnamefont {M.~P.}\ \bibnamefont
  {Howard}}, \bibinfo {author} {\bibfnamefont {A.~Z.}\ \bibnamefont
  {Panagiotopoulos}}, \ and\ \bibinfo {author} {\bibfnamefont {A.}~\bibnamefont
  {Nikoubashman}},\ }\href@noop {} {\bibfield  {journal} {\bibinfo  {journal}
  {J. Chem. Phys.}\ }\textbf {\bibinfo {volume} {142}},\ \bibinfo {pages}
  {224908} (\bibinfo {year} {2015})}\BibitemShut {NoStop}%
\bibitem [{\citenamefont {Trofa}\ \emph {et~al.}(2015)\citenamefont {Trofa},
  \citenamefont {Vocciante}, \citenamefont {D'Avino}, \citenamefont {Hulsen},
  \citenamefont {Greco},\ and\ \citenamefont
  {Maffettone}}]{trofa:compfluids:2015}%
  \BibitemOpen
  \bibfield  {author} {\bibinfo {author} {\bibfnamefont {M.}~\bibnamefont
  {Trofa}}, \bibinfo {author} {\bibfnamefont {M.}~\bibnamefont {Vocciante}},
  \bibinfo {author} {\bibfnamefont {G.}~\bibnamefont {D'Avino}}, \bibinfo
  {author} {\bibfnamefont {M.~A.}\ \bibnamefont {Hulsen}}, \bibinfo {author}
  {\bibfnamefont {F.}~\bibnamefont {Greco}}, \ and\ \bibinfo {author}
  {\bibfnamefont {P.~L.}\ \bibnamefont {Maffettone}},\ }\href@noop {}
  {\bibfield  {journal} {\bibinfo  {journal} {Comput. Fluids}\ }\textbf
  {\bibinfo {volume} {107}},\ \bibinfo {pages} {214} (\bibinfo {year}
  {2015})}\BibitemShut {NoStop}%
\bibitem [{\citenamefont {Prohm}, \citenamefont {Gierlak},\ and\ \citenamefont
  {Stark}(2012)}]{prohm:epje:12}%
  \BibitemOpen
  \bibfield  {author} {\bibinfo {author} {\bibfnamefont {C.}~\bibnamefont
  {Prohm}}, \bibinfo {author} {\bibfnamefont {M.}~\bibnamefont {Gierlak}}, \
  and\ \bibinfo {author} {\bibfnamefont {H.}~\bibnamefont {Stark}},\
  }\href@noop {} {\bibfield  {journal} {\bibinfo  {journal} {Eur. Phys. J. E}\
  }\textbf {\bibinfo {volume} {35}},\ \bibinfo {pages} {80} (\bibinfo {year}
  {2012})}\BibitemShut {NoStop}%
\bibitem [{\citenamefont {Mackay}\ \emph {et~al.}(2003)\citenamefont {Mackay},
  \citenamefont {Dao}, \citenamefont {Tuteja}, \citenamefont {Ho},
  \citenamefont {Van~Horn}, \citenamefont {Kim},\ and\ \citenamefont
  {Hawker}}]{Mackay:2003wq}%
  \BibitemOpen
  \bibfield  {author} {\bibinfo {author} {\bibfnamefont {M.~E.}\ \bibnamefont
  {Mackay}}, \bibinfo {author} {\bibfnamefont {T.~T.}\ \bibnamefont {Dao}},
  \bibinfo {author} {\bibfnamefont {A.}~\bibnamefont {Tuteja}}, \bibinfo
  {author} {\bibfnamefont {D.~L.}\ \bibnamefont {Ho}}, \bibinfo {author}
  {\bibfnamefont {B.}~\bibnamefont {Van~Horn}}, \bibinfo {author}
  {\bibfnamefont {H.-C.}\ \bibnamefont {Kim}}, \ and\ \bibinfo {author}
  {\bibfnamefont {C.~J.}\ \bibnamefont {Hawker}},\ }\href@noop {} {\bibfield
  {journal} {\bibinfo  {journal} {Nat. Mater.}\ }\textbf {\bibinfo {volume}
  {2}},\ \bibinfo {pages} {762} (\bibinfo {year} {2003})}\BibitemShut {NoStop}%
\bibitem [{\citenamefont {Wong}\ \emph {et~al.}(2004)\citenamefont {Wong},
  \citenamefont {Gardel}, \citenamefont {Reichman}, \citenamefont {Weeks},
  \citenamefont {Valentine}, \citenamefont {Bausch},\ and\ \citenamefont
  {Weitz}}]{Wong:2004cp}%
  \BibitemOpen
  \bibfield  {author} {\bibinfo {author} {\bibfnamefont {I.~Y.}\ \bibnamefont
  {Wong}}, \bibinfo {author} {\bibfnamefont {M.~L.}\ \bibnamefont {Gardel}},
  \bibinfo {author} {\bibfnamefont {D.~R.}\ \bibnamefont {Reichman}}, \bibinfo
  {author} {\bibfnamefont {E.~R.}\ \bibnamefont {Weeks}}, \bibinfo {author}
  {\bibfnamefont {M.~T.}\ \bibnamefont {Valentine}}, \bibinfo {author}
  {\bibfnamefont {A.~R.}\ \bibnamefont {Bausch}}, \ and\ \bibinfo {author}
  {\bibfnamefont {D.~A.}\ \bibnamefont {Weitz}},\ }\href@noop {} {\bibfield
  {journal} {\bibinfo  {journal} {Phys. Rev. Lett.}\ }\textbf {\bibinfo
  {volume} {92}},\ \bibinfo {pages} {178101} (\bibinfo {year}
  {2004})}\BibitemShut {NoStop}%
\bibitem [{\citenamefont {Tuteja}\ \emph {et~al.}(2007)\citenamefont {Tuteja},
  \citenamefont {Mackay}, \citenamefont {Narayanan}, \citenamefont {Asokan},\
  and\ \citenamefont {Wong}}]{Tuteja:2007dz}%
  \BibitemOpen
  \bibfield  {author} {\bibinfo {author} {\bibfnamefont {A.}~\bibnamefont
  {Tuteja}}, \bibinfo {author} {\bibfnamefont {M.~E.}\ \bibnamefont {Mackay}},
  \bibinfo {author} {\bibfnamefont {S.}~\bibnamefont {Narayanan}}, \bibinfo
  {author} {\bibfnamefont {S.}~\bibnamefont {Asokan}}, \ and\ \bibinfo {author}
  {\bibfnamefont {M.~S.}\ \bibnamefont {Wong}},\ }\href@noop {} {\bibfield
  {journal} {\bibinfo  {journal} {Nano Lett.}\ }\textbf {\bibinfo {volume}
  {7}},\ \bibinfo {pages} {1276} (\bibinfo {year} {2007})}\BibitemShut
  {NoStop}%
\bibitem [{\citenamefont {Poling-Skutvik}, \citenamefont {Krishnamoorti},\ and\
  \citenamefont {Conrad}(2015)}]{PolingSkutvik:2015be}%
  \BibitemOpen
  \bibfield  {author} {\bibinfo {author} {\bibfnamefont {R.}~\bibnamefont
  {Poling-Skutvik}}, \bibinfo {author} {\bibfnamefont {R.}~\bibnamefont
  {Krishnamoorti}}, \ and\ \bibinfo {author} {\bibfnamefont {J.~C.}\
  \bibnamefont {Conrad}},\ }\href@noop {} {\bibfield  {journal} {\bibinfo
  {journal} {ACS Macro Lett.}\ }\textbf {\bibinfo {volume} {4}},\ \bibinfo
  {pages} {1169} (\bibinfo {year} {2015})}\BibitemShut {NoStop}%
\bibitem [{\citenamefont {Chen}\ \emph {et~al.}(2018)\citenamefont {Chen},
  \citenamefont {Poling-Skutvik}, \citenamefont {Nikoubashman}, \citenamefont
  {Howard}, \citenamefont {Conrad},\ and\ \citenamefont
  {Palmer}}]{Chen:2018im}%
  \BibitemOpen
  \bibfield  {author} {\bibinfo {author} {\bibfnamefont {R.}~\bibnamefont
  {Chen}}, \bibinfo {author} {\bibfnamefont {R.}~\bibnamefont
  {Poling-Skutvik}}, \bibinfo {author} {\bibfnamefont {A.}~\bibnamefont
  {Nikoubashman}}, \bibinfo {author} {\bibfnamefont {M.~P.}\ \bibnamefont
  {Howard}}, \bibinfo {author} {\bibfnamefont {J.~C.}\ \bibnamefont {Conrad}},
  \ and\ \bibinfo {author} {\bibfnamefont {J.~C.}\ \bibnamefont {Palmer}},\
  }\href@noop {} {\bibfield  {journal} {\bibinfo  {journal} {Macromolecules}\
  }\textbf {\bibinfo {volume} {51}},\ \bibinfo {pages} {1865} (\bibinfo {year}
  {2018})}\BibitemShut {NoStop}%
\bibitem [{\citenamefont {Chen}\ \emph {et~al.}(2019)\citenamefont {Chen},
  \citenamefont {Poling-Skutvik}, \citenamefont {Howard}, \citenamefont
  {Nikoubashman}, \citenamefont {Egorov}, \citenamefont {Conrad},\ and\
  \citenamefont {Palmer}}]{Chen:2018:2}%
  \BibitemOpen
  \bibfield  {author} {\bibinfo {author} {\bibfnamefont {R.}~\bibnamefont
  {Chen}}, \bibinfo {author} {\bibfnamefont {R.}~\bibnamefont
  {Poling-Skutvik}}, \bibinfo {author} {\bibfnamefont {M.~P.}\ \bibnamefont
  {Howard}}, \bibinfo {author} {\bibfnamefont {A.}~\bibnamefont
  {Nikoubashman}}, \bibinfo {author} {\bibfnamefont {S.~A.}\ \bibnamefont
  {Egorov}}, \bibinfo {author} {\bibfnamefont {J.~C.}\ \bibnamefont {Conrad}},
  \ and\ \bibinfo {author} {\bibfnamefont {J.~C.}\ \bibnamefont {Palmer}},\
  }\href@noop {} {\bibfield  {journal} {\bibinfo  {journal} {Soft Matter}\ ,\
  \bibinfo {pages} {10.1039/C8SM01834K}} (\bibinfo {year} {2019})}\BibitemShut
  {NoStop}%
\bibitem [{\citenamefont {Hoogerbrugge}\ and\ \citenamefont
  {Koelman}(1992)}]{Hoogerbrugge:1992hl}%
  \BibitemOpen
  \bibfield  {author} {\bibinfo {author} {\bibfnamefont {P.~J.}\ \bibnamefont
  {Hoogerbrugge}}\ and\ \bibinfo {author} {\bibfnamefont {J.~M. V.~A.}\
  \bibnamefont {Koelman}},\ }\href@noop {} {\bibfield  {journal} {\bibinfo
  {journal} {Europhys. Lett.}\ }\textbf {\bibinfo {volume} {19}},\ \bibinfo
  {pages} {155} (\bibinfo {year} {1992})}\BibitemShut {NoStop}%
\bibitem [{\citenamefont {Espa{\~n}ol}\ and\ \citenamefont
  {Warren}(1995)}]{Espanol:1995hf}%
  \BibitemOpen
  \bibfield  {author} {\bibinfo {author} {\bibfnamefont {P.}~\bibnamefont
  {Espa{\~n}ol}}\ and\ \bibinfo {author} {\bibfnamefont {P.}~\bibnamefont
  {Warren}},\ }\href@noop {} {\bibfield  {journal} {\bibinfo  {journal}
  {Europhys. Lett.}\ }\textbf {\bibinfo {volume} {30}},\ \bibinfo {pages} {191}
  (\bibinfo {year} {1995})}\BibitemShut {NoStop}%
\bibitem [{\citenamefont {Groot}\ and\ \citenamefont
  {Warren}(1997)}]{Groot:1997du}%
  \BibitemOpen
  \bibfield  {author} {\bibinfo {author} {\bibfnamefont {R.~D.}\ \bibnamefont
  {Groot}}\ and\ \bibinfo {author} {\bibfnamefont {P.~B.}\ \bibnamefont
  {Warren}},\ }\href@noop {} {\bibfield  {journal} {\bibinfo  {journal} {J.
  Chem. Phys.}\ }\textbf {\bibinfo {volume} {107}},\ \bibinfo {pages} {4423}
  (\bibinfo {year} {1997})}\BibitemShut {NoStop}%
\bibitem [{\citenamefont {Visser}, \citenamefont {Hoefsloot},\ and\
  \citenamefont {Iedema}(2006)}]{Visser:2006bk}%
  \BibitemOpen
  \bibfield  {author} {\bibinfo {author} {\bibfnamefont {D.~C.}\ \bibnamefont
  {Visser}}, \bibinfo {author} {\bibfnamefont {H.~C.~J.}\ \bibnamefont
  {Hoefsloot}}, \ and\ \bibinfo {author} {\bibfnamefont {P.~D.}\ \bibnamefont
  {Iedema}},\ }\href@noop {} {\bibfield  {journal} {\bibinfo  {journal} {J.
  Comput. Phys.}\ }\textbf {\bibinfo {volume} {214}},\ \bibinfo {pages} {491}
  (\bibinfo {year} {2006})}\BibitemShut {NoStop}%
\bibitem [{\citenamefont {Fan}\ \emph {et~al.}(2006)\citenamefont {Fan},
  \citenamefont {Phan-Thien}, \citenamefont {Chen}, \citenamefont {Wu},\ and\
  \citenamefont {Ng}}]{Fan:2006ex}%
  \BibitemOpen
  \bibfield  {author} {\bibinfo {author} {\bibfnamefont {X.}~\bibnamefont
  {Fan}}, \bibinfo {author} {\bibfnamefont {N.}~\bibnamefont {Phan-Thien}},
  \bibinfo {author} {\bibfnamefont {S.}~\bibnamefont {Chen}}, \bibinfo {author}
  {\bibfnamefont {X.}~\bibnamefont {Wu}}, \ and\ \bibinfo {author}
  {\bibfnamefont {T.~Y.}\ \bibnamefont {Ng}},\ }\href@noop {} {\bibfield
  {journal} {\bibinfo  {journal} {Phys. Fluids}\ }\textbf {\bibinfo {volume}
  {18}},\ \bibinfo {pages} {063102} (\bibinfo {year} {2006})}\BibitemShut
  {NoStop}%
\bibitem [{\citenamefont {Anderson}, \citenamefont {Lorenz},\ and\
  \citenamefont {Travesset}(2008)}]{Anderson:2008vg}%
  \BibitemOpen
  \bibfield  {author} {\bibinfo {author} {\bibfnamefont {J.~A.}\ \bibnamefont
  {Anderson}}, \bibinfo {author} {\bibfnamefont {C.~D.}\ \bibnamefont
  {Lorenz}}, \ and\ \bibinfo {author} {\bibfnamefont {A.}~\bibnamefont
  {Travesset}},\ }\href@noop {} {\bibfield  {journal} {\bibinfo  {journal} {J.
  Comput. Phys.}\ }\textbf {\bibinfo {volume} {227}},\ \bibinfo {pages} {5342}
  (\bibinfo {year} {2008})}\BibitemShut {NoStop}%
\bibitem [{\citenamefont {Glaser}\ \emph {et~al.}(2015)\citenamefont {Glaser},
  \citenamefont {Nguyen}, \citenamefont {Anderson}, \citenamefont {Lui},
  \citenamefont {Spiga}, \citenamefont {Millan}, \citenamefont {Morse},\ and\
  \citenamefont {Glotzer}}]{Glaser:2015cu}%
  \BibitemOpen
  \bibfield  {author} {\bibinfo {author} {\bibfnamefont {J.}~\bibnamefont
  {Glaser}}, \bibinfo {author} {\bibfnamefont {T.~D.}\ \bibnamefont {Nguyen}},
  \bibinfo {author} {\bibfnamefont {J.~A.}\ \bibnamefont {Anderson}}, \bibinfo
  {author} {\bibfnamefont {P.}~\bibnamefont {Lui}}, \bibinfo {author}
  {\bibfnamefont {F.}~\bibnamefont {Spiga}}, \bibinfo {author} {\bibfnamefont
  {J.~A.}\ \bibnamefont {Millan}}, \bibinfo {author} {\bibfnamefont {D.~C.}\
  \bibnamefont {Morse}}, \ and\ \bibinfo {author} {\bibfnamefont {S.~C.}\
  \bibnamefont {Glotzer}},\ }\href@noop {} {\bibfield  {journal} {\bibinfo
  {journal} {Comput. Phys. Commun.}\ }\textbf {\bibinfo {volume} {192}},\
  \bibinfo {pages} {97} (\bibinfo {year} {2015})}\BibitemShut {NoStop}%
\bibitem [{\citenamefont {Phillips}, \citenamefont {Anderson},\ and\
  \citenamefont {Glotzer}(2011)}]{Phillips:2011td}%
  \BibitemOpen
  \bibfield  {author} {\bibinfo {author} {\bibfnamefont {C.~L.}\ \bibnamefont
  {Phillips}}, \bibinfo {author} {\bibfnamefont {J.~A.}\ \bibnamefont
  {Anderson}}, \ and\ \bibinfo {author} {\bibfnamefont {S.~C.}\ \bibnamefont
  {Glotzer}},\ }\href@noop {} {\bibfield  {journal} {\bibinfo  {journal} {J.
  Comput. Phys.}\ }\textbf {\bibinfo {volume} {230}},\ \bibinfo {pages} {7191}
  (\bibinfo {year} {2011})}\BibitemShut {NoStop}%
\bibitem [{\citenamefont {Kirkwood}\ and\ \citenamefont
  {Buff}(1949)}]{Kirkwood:1949gf}%
  \BibitemOpen
  \bibfield  {author} {\bibinfo {author} {\bibfnamefont {J.~G.}\ \bibnamefont
  {Kirkwood}}\ and\ \bibinfo {author} {\bibfnamefont {F.~P.}\ \bibnamefont
  {Buff}},\ }\href@noop {} {\bibfield  {journal} {\bibinfo  {journal} {J. Chem.
  Phys.}\ }\textbf {\bibinfo {volume} {17}},\ \bibinfo {pages} {338} (\bibinfo
  {year} {1949})}\BibitemShut {NoStop}%
\bibitem [{\citenamefont {Frenkel}\ and\ \citenamefont {Smit}(2002)}]{frenkel}%
  \BibitemOpen
  \bibfield  {author} {\bibinfo {author} {\bibfnamefont {D.}~\bibnamefont
  {Frenkel}}\ and\ \bibinfo {author} {\bibfnamefont {B.}~\bibnamefont {Smit}},\
  }\href@noop {} {\emph {\bibinfo {title} {{Understanding Molecular Simulation:
  From Algorithms to Applications}}}},\ \bibinfo {edition} {2nd}\ ed.\
  (\bibinfo  {publisher} {Academic Press},\ \bibinfo {year} {2002})\ p.\
  \bibinfo {pages} {472}\BibitemShut {NoStop}%
\bibitem [{\citenamefont {M{\"u}ller-Plathe}(1999)}]{MullerPlathe:1999vc}%
  \BibitemOpen
  \bibfield  {author} {\bibinfo {author} {\bibfnamefont {F.}~\bibnamefont
  {M{\"u}ller-Plathe}},\ }\href@noop {} {\bibfield  {journal} {\bibinfo
  {journal} {Phys. Rev. E}\ }\textbf {\bibinfo {volume} {59}},\ \bibinfo
  {pages} {4894} (\bibinfo {year} {1999})}\BibitemShut {NoStop}%
\bibitem [{\citenamefont {Statt}, \citenamefont {Howard},\ and\ \citenamefont
  {Panagiotopoulos}(2018)}]{Statt:2018}%
  \BibitemOpen
  \bibfield  {author} {\bibinfo {author} {\bibfnamefont {A.}~\bibnamefont
  {Statt}}, \bibinfo {author} {\bibfnamefont {M.~P.}\ \bibnamefont {Howard}}, \
  and\ \bibinfo {author} {\bibfnamefont {A.~Z.}\ \bibnamefont
  {Panagiotopoulos}},\ }\href@noop {} {\ ,\ \bibinfo {pages} {arXiv:1811.04097}
  (\bibinfo {year} {2018})}\BibitemShut {NoStop}%
\bibitem [{\citenamefont {Kranenburg}, \citenamefont {Nicolas},\ and\
  \citenamefont {Smit}(2004)}]{Kranenburg:2004fu}%
  \BibitemOpen
  \bibfield  {author} {\bibinfo {author} {\bibfnamefont {M.}~\bibnamefont
  {Kranenburg}}, \bibinfo {author} {\bibfnamefont {J.-P.}\ \bibnamefont
  {Nicolas}}, \ and\ \bibinfo {author} {\bibfnamefont {B.}~\bibnamefont
  {Smit}},\ }\href@noop {} {\bibfield  {journal} {\bibinfo  {journal} {Phys.
  Chem. Chem. Phys.}\ }\textbf {\bibinfo {volume} {6}},\ \bibinfo {pages}
  {4142} (\bibinfo {year} {2004})}\BibitemShut {NoStop}%
\bibitem [{\citenamefont {Pivkin}\ and\ \citenamefont
  {Karniadakis}(2005)}]{Pivkin:2005bu}%
  \BibitemOpen
  \bibfield  {author} {\bibinfo {author} {\bibfnamefont {I.~V.}\ \bibnamefont
  {Pivkin}}\ and\ \bibinfo {author} {\bibfnamefont {G.~E.}\ \bibnamefont
  {Karniadakis}},\ }\href@noop {} {\bibfield  {journal} {\bibinfo  {journal}
  {J. Comput. Phys.}\ }\textbf {\bibinfo {volume} {207}},\ \bibinfo {pages}
  {114} (\bibinfo {year} {2005})}\BibitemShut {NoStop}%
\bibitem [{\citenamefont {Fedosov}, \citenamefont {Pivkin},\ and\ \citenamefont
  {Karniadakis}(2008)}]{Fedosov:2008ct}%
  \BibitemOpen
  \bibfield  {author} {\bibinfo {author} {\bibfnamefont {D.~A.}\ \bibnamefont
  {Fedosov}}, \bibinfo {author} {\bibfnamefont {I.~V.}\ \bibnamefont {Pivkin}},
  \ and\ \bibinfo {author} {\bibfnamefont {G.~E.}\ \bibnamefont
  {Karniadakis}},\ }\href@noop {} {\bibfield  {journal} {\bibinfo  {journal}
  {J. Comput. Phys.}\ }\textbf {\bibinfo {volume} {227}},\ \bibinfo {pages}
  {2540} (\bibinfo {year} {2008})}\BibitemShut {NoStop}%
\bibitem [{\citenamefont {Revenga}, \citenamefont {Z{\'u}{\~n}iga},\ and\
  \citenamefont {Espa{\~n}ol}(1999)}]{Revenga:1999wl}%
  \BibitemOpen
  \bibfield  {author} {\bibinfo {author} {\bibfnamefont {M.}~\bibnamefont
  {Revenga}}, \bibinfo {author} {\bibfnamefont {I.}~\bibnamefont
  {Z{\'u}{\~n}iga}}, \ and\ \bibinfo {author} {\bibfnamefont {P.}~\bibnamefont
  {Espa{\~n}ol}},\ }\href@noop {} {\bibfield  {journal} {\bibinfo  {journal}
  {Comput. Phys. Commun.}\ }\textbf {\bibinfo {volume} {121--122}},\ \bibinfo
  {pages} {309} (\bibinfo {year} {1999})}\BibitemShut {NoStop}%
\bibitem [{\citenamefont {Stukowski}(2010)}]{Stukowski:2010ky}%
  \BibitemOpen
  \bibfield  {author} {\bibinfo {author} {\bibfnamefont {A.}~\bibnamefont
  {Stukowski}},\ }\href@noop {} {\bibfield  {journal} {\bibinfo  {journal}
  {Modelling Sim. Mater. Sci. Eng.}\ }\textbf {\bibinfo {volume} {18}},\
  \bibinfo {pages} {015012} (\bibinfo {year} {2010})}\BibitemShut {NoStop}%
\bibitem [{\citenamefont {Adorf}\ \emph {et~al.}(2018)\citenamefont {Adorf},
  \citenamefont {Dodd}, \citenamefont {Ramasubramani},\ and\ \citenamefont
  {Glotzer}}]{signac_commat}%
  \BibitemOpen
  \bibfield  {author} {\bibinfo {author} {\bibfnamefont {C.~S.}\ \bibnamefont
  {Adorf}}, \bibinfo {author} {\bibfnamefont {P.~M.}\ \bibnamefont {Dodd}},
  \bibinfo {author} {\bibfnamefont {V.}~\bibnamefont {Ramasubramani}}, \ and\
  \bibinfo {author} {\bibfnamefont {S.~C.}\ \bibnamefont {Glotzer}},\ }\href
  {\doibase 10.1016/j.commatsci.2018.01.035} {\bibfield  {journal} {\bibinfo
  {journal} {Comput. Mater. Sci.}\ }\textbf {\bibinfo {volume} {146}},\
  \bibinfo {pages} {220} (\bibinfo {year} {2018})}\BibitemShut {NoStop}%
\bibitem [{\citenamefont {Rubinstein}\ and\ \citenamefont
  {Colby}(2003)}]{Rubinstein}%
  \BibitemOpen
  \bibfield  {author} {\bibinfo {author} {\bibfnamefont {M.}~\bibnamefont
  {Rubinstein}}\ and\ \bibinfo {author} {\bibfnamefont {R.~H.}\ \bibnamefont
  {Colby}},\ }\href@noop {} {\emph {\bibinfo {title} {{Polymer Physics}}}}\
  (\bibinfo  {publisher} {Oxford University Press},\ \bibinfo {year} {2003})\
  pp.\ \bibinfo {pages} {314--318}\BibitemShut {NoStop}%
\bibitem [{\citenamefont {Doi}\ and\ \citenamefont {Edwards}(1986)}]{doi}%
  \BibitemOpen
  \bibfield  {author} {\bibinfo {author} {\bibfnamefont {M.}~\bibnamefont
  {Doi}}\ and\ \bibinfo {author} {\bibfnamefont {S.~F.}\ \bibnamefont
  {Edwards}},\ }\href@noop {} {\emph {\bibinfo {title} {{The Theory of Polymer
  Dynamics}}}}\ (\bibinfo  {publisher} {Clarendon Press},\ \bibinfo {address}
  {Oxford},\ \bibinfo {year} {1986})\ pp.\ \bibinfo {pages}
  {97--103}\BibitemShut {NoStop}%
\bibitem [{\citenamefont {Pedregosa}\ \emph {et~al.}(2011)\citenamefont
  {Pedregosa}, \citenamefont {Varoquaux}, \citenamefont {Gramfort},
  \citenamefont {Michel}, \citenamefont {Thirion}, \citenamefont {Grisel},
  \citenamefont {Blondel}, \citenamefont {Prettenhofer}, \citenamefont {Weiss},
  \citenamefont {Dubourg}, \citenamefont {Vanderplas}, \citenamefont {Passos},
  \citenamefont {Cournapeau}, \citenamefont {Brucher}, \citenamefont {Perrot},\
  and\ \citenamefont {Duchesnay}}]{scikit-learn}%
  \BibitemOpen
  \bibfield  {author} {\bibinfo {author} {\bibfnamefont {F.}~\bibnamefont
  {Pedregosa}}, \bibinfo {author} {\bibfnamefont {G.}~\bibnamefont
  {Varoquaux}}, \bibinfo {author} {\bibfnamefont {A.}~\bibnamefont {Gramfort}},
  \bibinfo {author} {\bibfnamefont {V.}~\bibnamefont {Michel}}, \bibinfo
  {author} {\bibfnamefont {B.}~\bibnamefont {Thirion}}, \bibinfo {author}
  {\bibfnamefont {O.}~\bibnamefont {Grisel}}, \bibinfo {author} {\bibfnamefont
  {M.}~\bibnamefont {Blondel}}, \bibinfo {author} {\bibfnamefont
  {P.}~\bibnamefont {Prettenhofer}}, \bibinfo {author} {\bibfnamefont
  {R.}~\bibnamefont {Weiss}}, \bibinfo {author} {\bibfnamefont
  {V.}~\bibnamefont {Dubourg}}, \bibinfo {author} {\bibfnamefont
  {J.}~\bibnamefont {Vanderplas}}, \bibinfo {author} {\bibfnamefont
  {A.}~\bibnamefont {Passos}}, \bibinfo {author} {\bibfnamefont
  {D.}~\bibnamefont {Cournapeau}}, \bibinfo {author} {\bibfnamefont
  {M.}~\bibnamefont {Brucher}}, \bibinfo {author} {\bibfnamefont
  {M.}~\bibnamefont {Perrot}}, \ and\ \bibinfo {author} {\bibfnamefont
  {E.}~\bibnamefont {Duchesnay}},\ }\href@noop {} {\bibfield  {journal}
  {\bibinfo  {journal} {J. Mach. Learn. Res.}\ }\textbf {\bibinfo {volume}
  {12}},\ \bibinfo {pages} {2825} (\bibinfo {year} {2011})}\BibitemShut
  {NoStop}%
\bibitem [{\citenamefont {Ester}\ \emph {et~al.}(1999)\citenamefont {Ester},
  \citenamefont {{Kriegel, Hans-Peter}}, \citenamefont {Sander},\ and\
  \citenamefont {Xu}}]{Ester:1999tm}%
  \BibitemOpen
  \bibfield  {author} {\bibinfo {author} {\bibfnamefont {M.}~\bibnamefont
  {Ester}}, \bibinfo {author} {\bibnamefont {{Kriegel, Hans-Peter}}}, \bibinfo
  {author} {\bibfnamefont {J.}~\bibnamefont {Sander}}, \ and\ \bibinfo {author}
  {\bibfnamefont {X.}~\bibnamefont {Xu}},\ }in\ \href@noop {} {\emph {\bibinfo
  {booktitle} {Proceedings of the 2nd International Conference on Knowledge
  Discovery and Data Mining}}}\ (\bibinfo  {publisher} {AAAI Press},\ \bibinfo
  {address} {Portland, OR},\ \bibinfo {year} {1999})\ pp.\ \bibinfo {pages}
  {226--231}\BibitemShut {NoStop}%
\bibitem [{\citenamefont {Asakura}\ and\ \citenamefont
  {Oosawa}(1954)}]{Asakura:1954ts}%
  \BibitemOpen
  \bibfield  {author} {\bibinfo {author} {\bibfnamefont {S.}~\bibnamefont
  {Asakura}}\ and\ \bibinfo {author} {\bibfnamefont {F.}~\bibnamefont
  {Oosawa}},\ }\href@noop {} {\bibfield  {journal} {\bibinfo  {journal} {J.
  Chem. Phys.}\ }\textbf {\bibinfo {volume} {22}},\ \bibinfo {pages} {1255}
  (\bibinfo {year} {1954})}\BibitemShut {NoStop}%
\bibitem [{\citenamefont {Russel}, \citenamefont {Saville},\ and\ \citenamefont
  {Schowalter}(1989)}]{Russel:1989}%
  \BibitemOpen
  \bibfield  {author} {\bibinfo {author} {\bibfnamefont {W.~B.}\ \bibnamefont
  {Russel}}, \bibinfo {author} {\bibfnamefont {D.~A.}\ \bibnamefont {Saville}},
  \ and\ \bibinfo {author} {\bibfnamefont {W.~R.}\ \bibnamefont {Schowalter}},\
  }\href@noop {} {\emph {\bibinfo {title} {{Colloidal Dispersions}}}}\
  (\bibinfo  {publisher} {Cambridge University Press},\ \bibinfo {year}
  {1989})\BibitemShut {NoStop}%
\bibitem [{\citenamefont {Taylor}(1932)}]{Taylor:1932tm}%
  \BibitemOpen
  \bibfield  {author} {\bibinfo {author} {\bibfnamefont {G.~I.}\ \bibnamefont
  {Taylor}},\ }\href@noop {} {\bibfield  {journal} {\bibinfo  {journal} {Proc.
  R. Soc. London A}\ }\textbf {\bibinfo {volume} {138}},\ \bibinfo {pages} {41}
  (\bibinfo {year} {1932})}\BibitemShut {NoStop}%
\bibitem [{\citenamefont {Taylor}(1934)}]{Taylor:1934wx}%
  \BibitemOpen
  \bibfield  {author} {\bibinfo {author} {\bibfnamefont {G.~I.}\ \bibnamefont
  {Taylor}},\ }\href@noop {} {\bibfield  {journal} {\bibinfo  {journal} {Proc.
  R. Soc. London A}\ }\textbf {\bibinfo {volume} {146}},\ \bibinfo {pages}
  {501} (\bibinfo {year} {1934})}\BibitemShut {NoStop}%
\bibitem [{\citenamefont {Ripoll}, \citenamefont {Winkler},\ and\ \citenamefont
  {Gompper}(2006)}]{Ripoll:2006dh}%
  \BibitemOpen
  \bibfield  {author} {\bibinfo {author} {\bibfnamefont {M.}~\bibnamefont
  {Ripoll}}, \bibinfo {author} {\bibfnamefont {R.~G.}\ \bibnamefont {Winkler}},
  \ and\ \bibinfo {author} {\bibfnamefont {G.}~\bibnamefont {Gompper}},\
  }\href@noop {} {\bibfield  {journal} {\bibinfo  {journal} {Phys. Rev. Lett.}\
  }\textbf {\bibinfo {volume} {96}},\ \bibinfo {pages} {188302} (\bibinfo
  {year} {2006})}\BibitemShut {NoStop}%
\bibitem [{\citenamefont {Rallison}\ and\ \citenamefont
  {Acrivos}(1978)}]{Rallison:1978dd}%
  \BibitemOpen
  \bibfield  {author} {\bibinfo {author} {\bibfnamefont {J.~M.}\ \bibnamefont
  {Rallison}}\ and\ \bibinfo {author} {\bibfnamefont {A.}~\bibnamefont
  {Acrivos}},\ }\href@noop {} {\bibfield  {journal} {\bibinfo  {journal} {J.
  Fluid Mech.}\ }\textbf {\bibinfo {volume} {89}},\ \bibinfo {pages} {191}
  (\bibinfo {year} {1978})}\BibitemShut {NoStop}%
\bibitem [{\citenamefont {Pan}, \citenamefont {Phan-Thien},\ and\ \citenamefont
  {Khoo}(2014)}]{Pan:2014bp}%
  \BibitemOpen
  \bibfield  {author} {\bibinfo {author} {\bibfnamefont {D.}~\bibnamefont
  {Pan}}, \bibinfo {author} {\bibfnamefont {N.}~\bibnamefont {Phan-Thien}}, \
  and\ \bibinfo {author} {\bibfnamefont {B.~C.}\ \bibnamefont {Khoo}},\
  }\href@noop {} {\bibfield  {journal} {\bibinfo  {journal} {J. Non-Newtonian
  Fluid Mech.}\ }\textbf {\bibinfo {volume} {212}},\ \bibinfo {pages} {63}
  (\bibinfo {year} {2014})}\BibitemShut {NoStop}%
\bibitem [{\citenamefont {Srivastva}\ and\ \citenamefont
  {Nikoubashman}(2018)}]{Srivastva:2018ft}%
  \BibitemOpen
  \bibfield  {author} {\bibinfo {author} {\bibfnamefont {D.}~\bibnamefont
  {Srivastva}}\ and\ \bibinfo {author} {\bibfnamefont {A.}~\bibnamefont
  {Nikoubashman}},\ }\href@noop {} {\bibfield  {journal} {\bibinfo  {journal}
  {Polymers}\ }\textbf {\bibinfo {volume} {10}},\ \bibinfo {pages} {599}
  (\bibinfo {year} {2018})}\BibitemShut {NoStop}%
\end{thebibliography}%

\end{document}